\documentclass[%
 reprint,
 twocolumn,
 amsmath,amssymb,
aip,jcp,
]{revtex4-2}

\usepackage{natbib}
\usepackage{graphicx}
\usepackage{dcolumn}
\usepackage{bm}
\usepackage{xcolor}
\usepackage{physics}
\usepackage{lipsum}
\usepackage{placeins}

\begin{document}
\title{Nonadiabatic to Adiabatic Transition of Electron Transfer in Colloidal Quantum Dot Molecules}
\author{Bokang Hou}
\address{Department of Chemistry, University of
California, Berkeley, California 94720, United States;}
\author{Michael Thoss}
\address{Institute of Physics, Albert-Ludwigs University Freiburg, Hermann-Herder-Straße 3, 79104 Freiburg, Germany;}
\author{Uri Banin}
\address{Institute of Chemistry and the Center for
Nanoscience and Nanotechnology, The Hebrew University of
Jerusalem, 91904 Jerusalem, Israel}
\author{Eran Rabani}
\address{Department of Chemistry, University of
California, Berkeley, California 94720, United States;}
\address{Materials Sciences Division, Lawrence Berkeley National
Laboratory, Berkeley, California 94720, United States;}
\address{The
Raymond and Beverly Sackler Center of Computational
Molecular and Materials Science, Tel Aviv University, Tel
Aviv 69978, Israel;}

\begin{abstract}
Electron transfer is an important and fundamental process in chemistry, biology and physics, and has received significant attention in recent years. Perhaps one of the most intriguing questions concerns with the realization of the transitions between nonadiabatic and adiabatic regimes of electron transfer, as the coupling (hybridization) energy, $J$, between the donor and acceptor is varied. Here, using colloidal quantum dot molecules, a new class of coupled quantum dot dimers, we computationally demonstrate how the hybridization energy between the donor and acceptor quantum dots can be tuned by simply changing the neck dimensions and/or the quantum dot size. This provides a handle to tune the electron transfer from the nonadiabatic over-damped Marcus regime to the coherent adiabatic regime in a single system, without changing the reorganization energy, $\lambda$, or the typical phonon frequency, $\omega_c$. We develop an atomistic model to account for several donor and acceptor states and how they couple to the lattice vibrations, and utilize the Ehrenfest mean-field mixed quantum-classical method to describe the charge transfer dynamics as the nonadiabatic parameter, $\gamma$, is varied. We find that charge transfer rates increase by several orders of magnitude as the system is driven to the coherent, adiabatic limit, even at elevated temperatures, and delineate the inter-dot and torsional acoustic modes that couple most strongly to the charge transfer reaction coordinate. 
\end{abstract}
\maketitle

\begin{figure}[t]
    \centering
    {{\includegraphics[width=8.0cm]{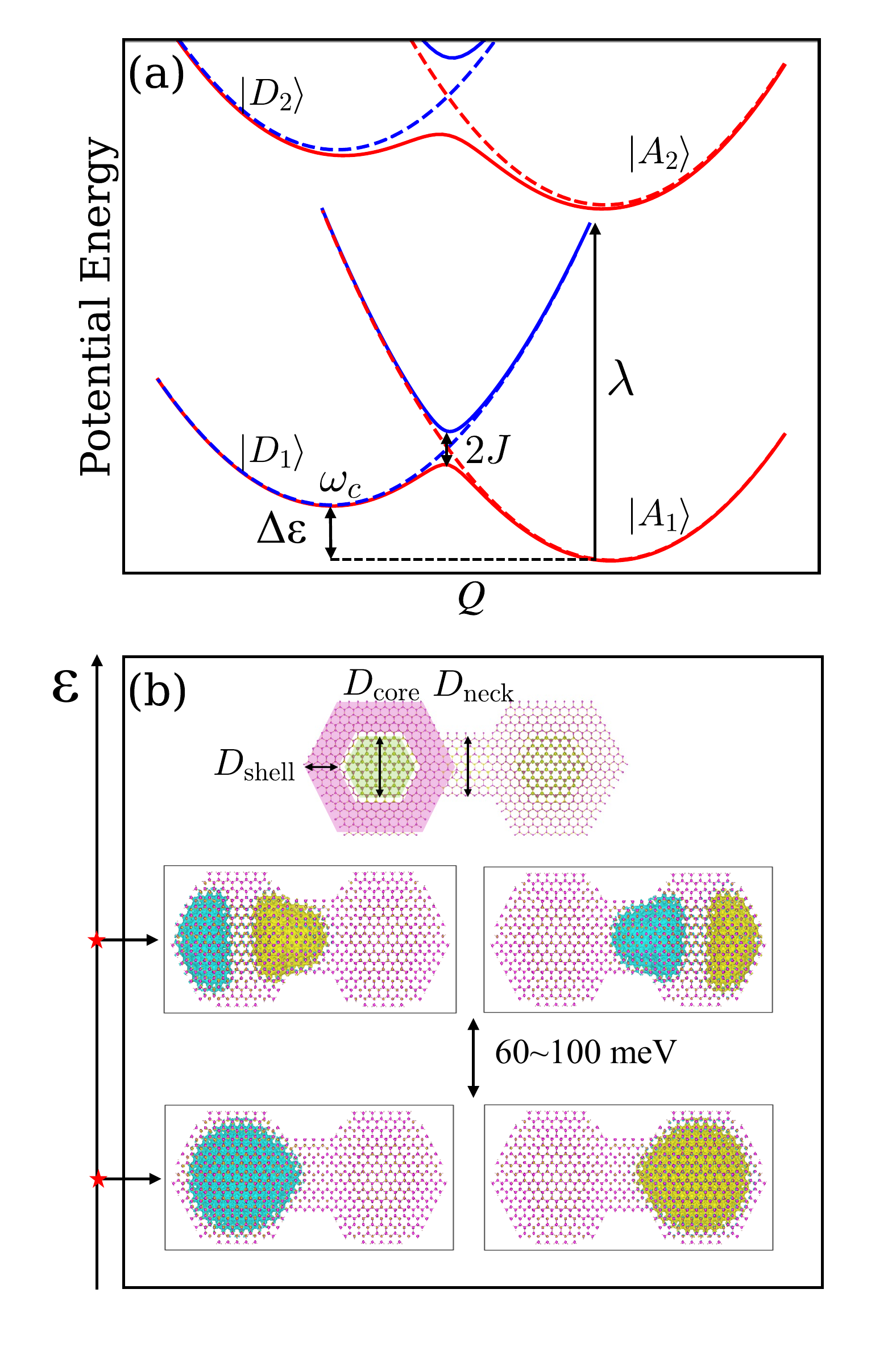} }}
    \caption{(a) Schematic sketch of the potential energy surfaces along the reaction coordinate~\cite{marcus_theory_1956} $Q$ and the energy scales appearing in Eq.~(\ref{eq:marcus}). The solid and dashed lines represent the adiabatic and diabatic potential energy surfaces, respectively. (b) Hyper-sphere plots of donor (left NC) and acceptor states (right NC) obtained from the semi-empirical pseudopotential calculation and FB-localization of the quasi-electron eigenstates in wurtzite CQD dimers. The energy difference between $1S_{\rm e}$-like ($\ket{D_1}$/$\ket{A_1}$) and $1P_{\rm e}$-like ($\ket{D_2}$/$\ket{A_2}$) orbitals is in the range of $60\sim100$ meV, depending on the size of the NC building blocks.}
    \label{fig:ele_struct}
\end{figure}

The theoretical framework to describe charge transfer reactions between a donor and an acceptor dates back to the seminal work of Marcus.\cite{marcus_theory_1956} Using a semi-classical perturbative approach, Marcus derived a relation for the outer-shell electron transfer rate at high temperature ($T$) in terms of the driving force ($\Delta \varepsilon$), the reorganization energy ($\lambda$) characterizing the coupling to nuclear fluctuations, and the hybridization/electronic couplings between donor and acceptor states ($J$), assumed to be small (see Fig.~\ref{fig:ele_struct}(a) for a sketch of the relevant energy scales):
\begin{equation}
\label{eq:marcus}
k_{\rm M}=\frac{\left|J\right|^2}{\hbar} \sqrt{\frac{\pi}{\lambda k_{\rm B} T}} \exp{\frac{-\left(\Delta\varepsilon+\lambda\right)^2}{4 \lambda k_{\rm B} T}}.
\end{equation}
Marcus theory is suitable for the {\em nonadiabatic} electron transfer regime, i.e. for $\gamma \ll 1$, where $\gamma$ is the adiabatic parameter~\cite{newton_electron_1984} defined by the ratio of the donor-acceptor coupling, the characteristic phonon bath frequency ($\omega_c$), and the reorganization energy
\begin{equation}\label{eq:gamma}
\gamma=\frac{|J|^2}{2\hbar \omega_c} \sqrt{\frac{\pi}{\lambda k_{\rm B} T}}.
\end{equation}

Marcus nonadiabatic electron transfer theory was extended in several different directions. Jortner and coworkers described the role of quantum nuclear fluctuations as well as non-parabolicities in the donor and acceptor free energies on the electron transfer rate.\cite{jortner_temperature_1976} Redfield theory \cite{redfield_theory_1965} was used to extend Marcus theory to account for coherences between the donor and acceptor, observed in photosynthetic complexes and quantum dot islands exhibiting strong electronic couplings.\cite{engel_environment-assisted_2014,heinrich_quantum-coherent_2021} Unlike the Marcus regime in which the transfer dynamics decay exponentially and can be characterized by a rate constant, the population dynamics in this coherent limit oscillate between the donor and acceptor, eventually relaxing to equilibrium. Zusman developed a framework to describe the crossover from the Marcus weak hybridization limit to the {\em adiabatic} limit.\cite{zusman_outer-sphere_1980,zusman_theory_1983}  In the adiabatic limit, the pre-exponential factor appearing in Marcus' rate theory (c.f., Eq.~\eqref{eq:marcus}) becomes independent of the electron tunneling probability between the donor and the acceptor, and the rate constant is given by transition state rate theory:
\begin{equation}
\label{eq:zusman}
k_{\rm TST}=\frac{\omega_c}{4} \sqrt{\frac{\lambda}{\pi k_{\rm B} T}} \exp{\frac{-\left(\Delta \varepsilon+\lambda\right)^2}{4 \lambda k_{\rm B} T}}.
\end{equation}

Experimental manifestation of transition from nonadiabatic to adiabatic charge transfer is challenging and requires exquisite control over electronic and vibrational degrees of freedom.\cite{demadis_localized--delocalized_2001,fiebig_femtosecond_2001} Recently, Zhu \textit{et al.}\cite{zhu_crossover_2021} studied electron transfer reactions in mixed valance donor-bridge-acceptor complexes. By changing the length of the bridge connecting the donor and acceptor and the functional groups on both the donor and the acceptor, they were able to drive the system from the adiabatic to the nonadiabatic regime. However, their approach was limited to a either $1$, $2$, or $3$ bridge units, covering a narrow range of behaviors. 

In this work, we revisit this problem and consider the charge transfer between two coupled colloidal quantum dot (CQD) nanocrystals (NCs) that are connected by a neck/bridge (see  Fig.~\ref{fig:ele_struct}(b)). Charge transfer in such systems is particularly interesting due to the flexibility in designing the donor and acceptor states by, for example, changing the width of the neck ($D_\text{neck}$) between two NCs and/or the diameter of each NC cores ($D_\text{core}$), as well as controlling the shell thickness ($D_\text{shell}$).\cite{cui_colloidal_2019,cui_neck_2021} By continuously varying  these parameters, the hybridization between the donor and the acceptor, $J$, can be tuned across a wide range of values while at the same time the reorganization energy, $\lambda$, and the typical vibrations frequency, $\omega_c$, change slightly. Furthermore, the hybridization energies can be tuned to be larger than the thermal energy at room temperature. This is quite distinct from the behavior of molecular junction, where a change in the hybridization energy is often accompanied by a change to the reorganization energy and the vibrational frequency. This is because control over these parameters in molecular junctions is achieved by either changing the donor/acceptor molecules, or by extending the bridge connecting them, which result in changes in the other parameters as well.

From a theoretical/computational perspective, this system poses several challenges, particularly with respect to the dimensions and number of valance electrons. Therefore, our approach to describe charge transfer is based on a model Hamiltonian which is parameterized by first-principle and semiempirical calculations. The total Hamiltonian can be divided into a sum of three terms, $H = H_{S} + H_{B} + H_{SB}$, where $H_{S}$ describes the electronic system (donor and acceptor states and their hybridization), $H_{B}$ is the Hamiltonian for the nuclear degrees of freedom (DOF) of the QD dimer (nuclear vibrations), and the interaction between the electronic system and the nuclear vibrations, approximated to the lowest order in the nuclear DOF,\cite{jasrasaria_interplay_2021} is described by $H_{SB}$. The three terms are given by:
\begin{equation}
\begin{aligned}
    &H_{S} = \sum_{n\in \mathcal{D}, \mathcal{A}} \varepsilon_{n}\ket{\phi_n}\bra{\phi_n} + \sum_{\substack{n\in \mathcal{D} \\ m\in \mathcal{A}}} J_{nm} \ket{\phi_n}\bra{\phi_m} + h.c.  \\
    &H_{B} = \sum_{\alpha}\frac{1}{2}{P}_{\alpha}^2 + U(Q_1, Q_2, \dots) \\
    &H_{SB} =\sum_{\alpha}\sum_{\substack{n,m\in \mathcal{D},\mathcal{A}}} \ket{\phi_n}\bra{\phi_m}V_{nm}^\alpha{Q}_{\alpha}.
\end{aligned}
\end{equation}
In the above equations, $\varepsilon_n$ is the energy of state $\ket{\phi_n}$ ($n \in \mathcal{D},\mathcal{A}$ with $\mathcal{D}=\{D_1,D_2\dots\}$, $\mathcal{A}=\{A_1,A_2\dots\}$) and $J_{nm}$ is the hybridization term between the donor state $\ket{\phi_n}$ and acceptor state $\ket{\phi_m}$. To obtain the donor and acceptor states, we used the semi-empirical pseudopotential model~\cite{wang_electronic_1994,rabani_electronic_1999} to describe the quasi-electron Hamiltonian ($\hat{h}_{\rm QP}$) and the filter diagonalization technique~\cite{wall_extraction_1995,toledo_very_2002} to calculate the eigenstates ($\ket{\psi_n}$) of the dimer near the bottom of the conduction band. The F\"{o}rster-Boys  localization scheme~\cite{foster_canonical_1960,kleier_localized_1974} was then used to generated the local donor and acceptor states ($\ket{\phi_n}$) from the eigenstates ($\ket{\psi_n}$), yielding $\varepsilon_n= \bra{\phi_n}\hat{h}_{\rm QP}\ket{\phi_n}$ and $J_{nm}=\bra{\phi_n}\hat{h}_{\rm QP}\ket{\phi_m}$ for $n\in\mathcal{D}, m\in\mathcal{A}$, and otherwise by construction, it is set to $J_{nm}=0$. We use the  Stilinger-Weber~\cite{zhou_stillinger-weber_2013} potential energy surface to describe the nuclear degrees of freedom, where $P_\alpha$ and $Q_\alpha$ are the $\alpha$ mass-weighted vibrational normal mode coordinates, determined by diagonlizing the Hessian matrix at the equilibrium geometry.\cite{jasrasaria_interplay_2021} The strength of coupling between states $\ket{\phi_n}$ and $\ket{\phi_m}$ to mode $\alpha$ is given by $V_{nm}^\alpha$, and is determined directly using the pseudopotential Hamiltonian\cite{jasrasaria_interplay_2021} (see Supporting Information Eq. (6) for more information). 

\begin{figure*}[t]
    \centering
    {{\includegraphics[width=15cm]{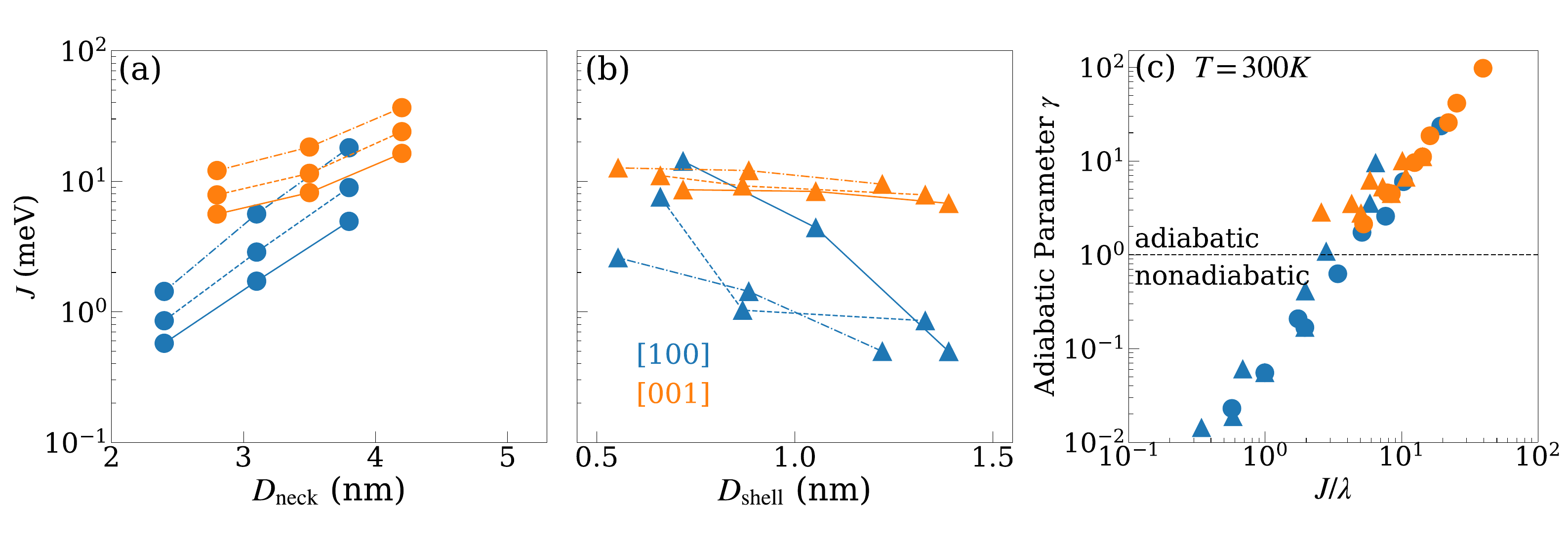} }}
    \caption{Hybridization energy $J$ of the ground donor and acceptor states as a function of (a) $D_\text{neck}$ and (b) $D_\text{shell}$ for [100] and [001] orientation attachments. Solid, dashed, and dotted-dashed lines correspond to $D_\text{core}=2.2$nm, $D_\text{core}=3.0$nm, $D_\text{core}=3.9$nm. (c) Adiabatic parameter $\gamma$ vs. $J/\lambda$ for varying neck-core sizes or shell-core sizes at 300 K. We define $\gamma=1$ as the crossover from the nonadiabatic to adiabatic electron transfer regimes. }
    \label{fig:J_gamma}
\end{figure*}

\begin{figure*}
\centering
    {{\includegraphics[width=15cm]{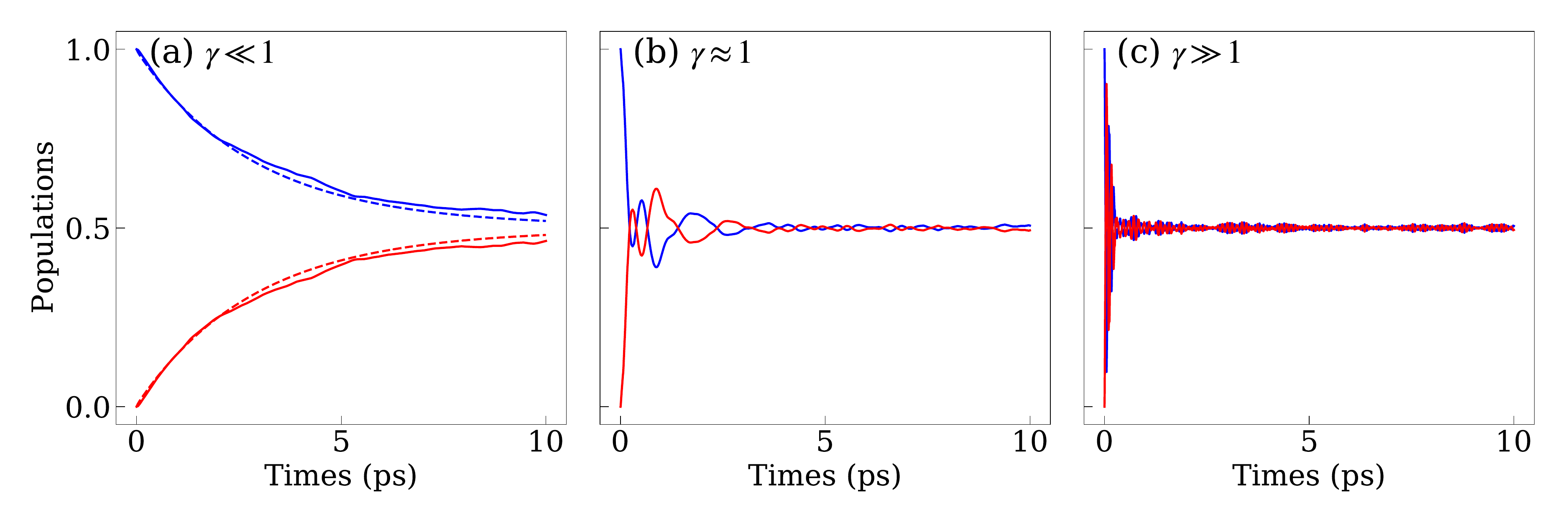} }}
    \caption{Population dynamics of the donor, $p_\mathcal{D}(t)$ (blue), and the acceptor, $p_\mathcal{A}(t)$ (red), at $T=300$K for three regimes: (a)~nonadiabatic, $D_{\rm neck}=2.4$ nm, $D_{\rm core}=2.2$ nm (b)~intermediate, $D_{\rm neck}=3.1$ nm, $D_{\rm core}=3.0$ nm (c)~adiabatic, $D_{\rm neck}=3.8$ nm, $D_{\rm core}=3.9$ nm. All structures are in symmetric attachment, leading to  $p_\mathcal{D}=p_\mathcal{A}=\frac{1}{2}$ in the long-time limit. The dashed lines in (a) are the population generated from master equation with rates computed using Marcus theory.}
    \label{fig:pop_panel}
\end{figure*}

Several low lying donor and acceptor states for the attachment orientation $[100]$ are shown in Fig.~\ref{fig:ele_struct}(b). The $[100]$ attachment results in a symmetric distribution of the charge density, similar to a homo-nuclear diatomic molecule, with an "atomic-like" basis of an effective mass particle-in-a-sphere  The two lowest energy donor/acceptor states are mainly comprised of $1S_{\rm e}$-like orbitals, while higher lying states show $1P_{\rm e}$-like character (either along or perpendicular to the dimer axis). For the systems considered in this work, the energy gap between $1S_{\rm e}$-like and $1P_{\rm e}$-like orbitals ranges from $60$ to $100$~meV. The symmetry is broken along the $[001]$ orientation attachment, resulting in a small energy bias and an asymmetric charge distribution, consistent with the behavior of hetero-nuclear diatomic molecules (see Fig.~S1 for more information).

In Fig.~\ref{fig:J_gamma} panels (a) and (b) we plot the hybridization energy between the ground donor and acceptor states, $J=J_{D_1A_1}$, as a function of the neck ($D_{\rm neck}$) and shell ($D_{\rm shell}$) widths, respectively, with different NC core diameters ($D_{\rm core}$). As expected, increasing the neck width and core diameters or decreasing the shell thickness, results in an exponential increase of the magnitude of the hybridization energy. Furthermore, the $[100]$ attachment (blue curves) shows a much steeper dependence on $D_{\rm neck}$ and $D_{\rm shell}$, as a result of the larger core-to-core distance for this orientation. In Fig.~\ref{fig:J_gamma}~(c) we plot the adiabatic parameter, $\gamma$, as a function of the hybridization energy for the same set of neck and shell dimensions shown in panels (a) and (b).  We find a crossover from the nonadiabatic to the adiabatic electron transfer regimes as $J \rightarrow  \lambda$. Since the characteristic vibrational frequency and the reorganization energy depend weakly on the dimer geometry, we find that the crossover mainly is affected by the hybridization energy. Note that for this set of building block monomers, the electron transfer in the $[001]$ orientation attachment is in the adiabatic limit regardless of the neck sizes or shell thickness, while it can be tuned from the nonadiabatic to the adiabatic limits for the $[100]$ orientation attachment. The dimensions of CQD dimers in Fig.~\ref{fig:J_gamma} are summarized in the Supporting Information, tables~S1 and S2. 

\begin{figure*}[t]
    \centering
    {{\includegraphics[width=12cm]{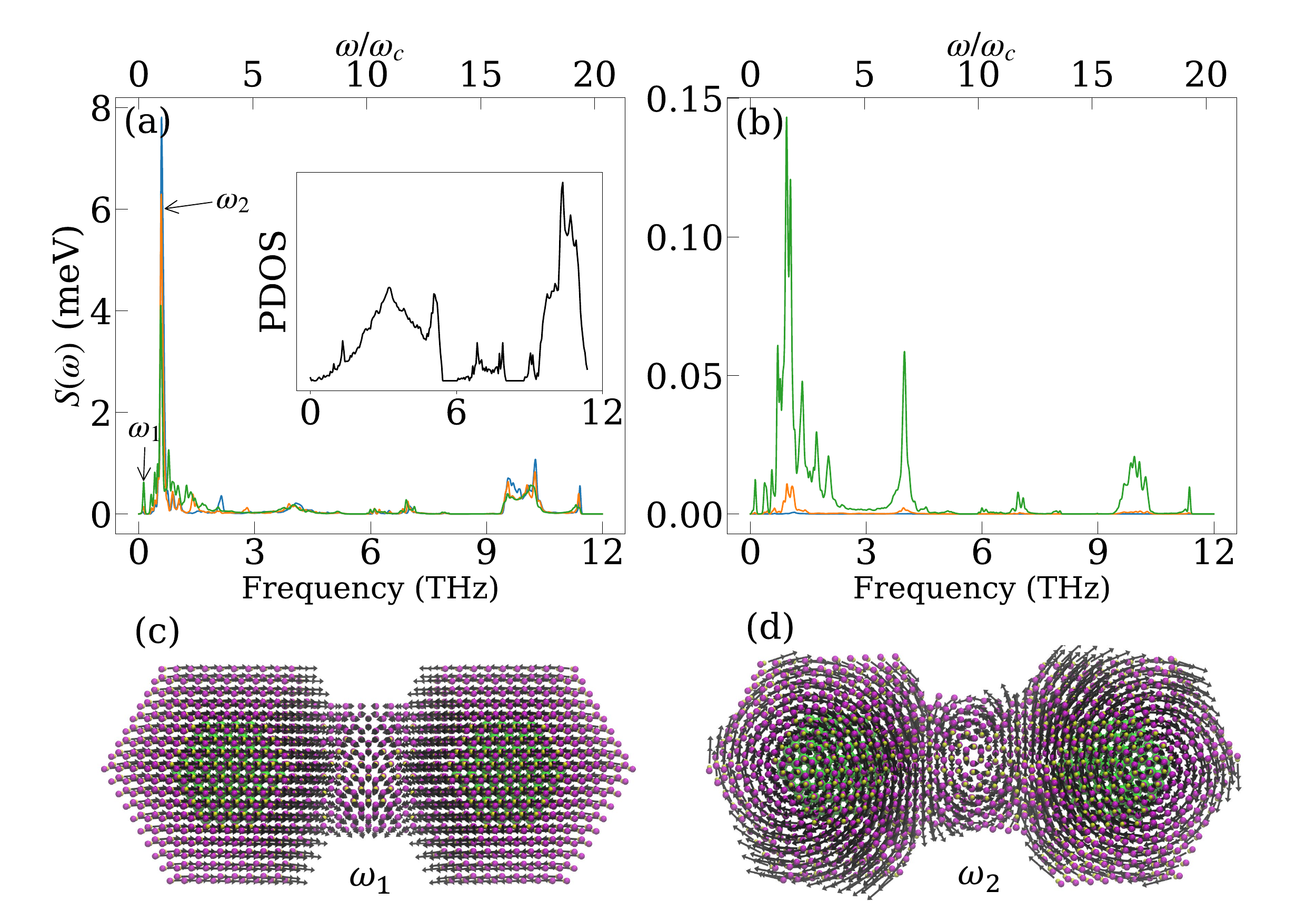} }}
    \caption{(a) Diagonal spectral densities $S_{D_1D_1}(\omega)$ and (b) off-diagonal spectral densities $S_{D_1A_1}(\omega)$ for the same systems studied in Fig.~\ref{fig:pop_panel}. The blue, orange and green lines correspond to Fig.~\ref{fig:pop_panel} panel (a), (b) and (c). The inset shows the phonon density of state (PDOS).  The two most important modes for contributing to dephasing are the acoustic inter-dot vibrational mode ($\omega_1 = 0.13$ THz) shown in panel (c) and the torsional ($\omega_2 = 0.58$ THz) mode shown in panel (d). The arrows indicate the motion of each atom for each mode.}
    \label{fig:vibronic}
\end{figure*}

In Fig.~\ref{fig:pop_panel} we plots the donor and acceptor populations as a function of time in the nonadiabatic-Marcus regime ($\gamma \ll 1$), the intermediate regime ($\gamma \approx 1$), and the adiabatic regime ($\gamma \gg 1$). We used the Ehrenfest mean-field mixed quantum-classical method~\cite{crespo-otero_recent_2018,mclachlan_variational_1964,nijjar_ehrenfest_2019} to describe the dynamics in all three regimes. As discussed below (c.f., Fig.~\ref{fig:vibronic}), the electron is mainly coupled to the low frequency acoustic modes, for which the classical limit is adequate ($\hbar \omega_c \ll k_{\rm B}T$). For $\gamma \ll 1$ in Fig.~\ref{fig:pop_panel} (a) we also compare the mean-field results to a master equation (due to the presence of multiple donor and acceptor states) with transition rates obtained from Marcus theory. The donor ($p_\mathcal{D}(t)$) and acceptor ($p_\mathcal{A}(t)$) populations are given as projections onto the donor and acceptor Hilbert spaces, respectively. Individual state populations corresponding to the results shown in Fig.~\ref{fig:pop_panel} are shown in the Supporting Information Fig.~S2 (population dynamics for all other structures studied in this work are shown in Fig.~S3 to S6).

For weak hybridization between the donor and acceptor (i.e., for small $D_{\rm neck}$ and/or $D_{\rm shell}$), the population dynamics are characterized by an over-damped exponential decay shown in Fig.~\ref{fig:pop_panel} (a), with a decay rate that approximately matches the Marcus rate between the ground donor and ground acceptor states. For intermediate values of the shell and neck thicknesses, i.e., for $\gamma \approx 1$, the population dynamics show underdamped coherent oscillations, with a Rabi frequency that matches predominately the ground state donor-acceptor hybridization (Fig.~\ref{fig:pop_panel}(b)). As the neck thickness is further increased and/or the shell thickness is further decreased, the Rabi period shortens, signifying the increase in the hybridization energy, as shown in Fig.~\ref{fig:pop_panel} (c). 

The agreement between the mean field theory and the perturbative Marcus result for $\gamma \ll 1$ is consistent with the so called "average classical limit",\cite{egorov_vibronic_1998} where the dynamics of the nuclear degrees of freedom are governed by the arithmetic average donor/acceptor Hamiltonian.\cite{egorov_nonradiative_1999} The average classical limit has been motivated by the analysis of the Wigner form of the quantum mechanical expression for the relevant time-correlation function,\cite{shemetulskis_semiclassical_1992} suggesting that the average Hamiltonian (similar to mean field) provides the most accurate approximation to the fully quantum mechanical results.\cite{egorov_nonradiative_1999} For large values of $\gamma$, where the hybridization energy is much larger than the reorganization energy ($J \gg \lambda$) and the system is in the weak electron-phonon coupling limit, the average mean force on the nuclei is similar to the force for each diabatic potential energy surfaces, and the mean field dynamics accurately reproduce the many-body solution.\footnote{Quantum mechanical test calculations using the multiconfiguration time-dependent Hartree (MCTDH) method show good agreement with the results of the Ehrenfest method. }

\begin{figure*}[t]
    \centering
    {{\includegraphics[width=16cm]{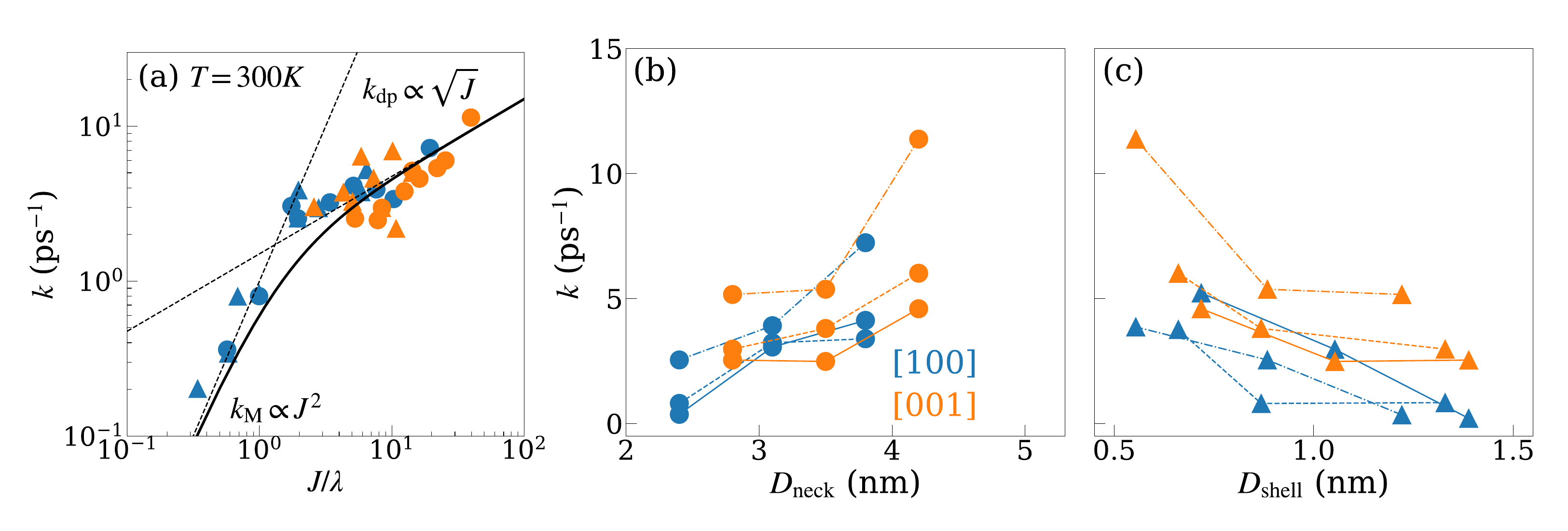} }}
    \caption{(a) Electron transfer rates, $k$, plotted as a function of $J/\lambda$ at $T=300$K for all structures studies in this work. The solid black curve is a fit to a connection formula, $k^{-1}=k_{\rm M}^{-1} + k_{\rm dp}^{-1}$\, with an empirical form for $k_{\rm dp}\propto\sqrt{J}$ shown as dashed line. We also show the $J\rightarrow 0$ Marcus limit, $k_{\rm M}^{-1}$. (b) and (c): Electron transfer rates plotted as a function of the neck and shell thicknesses, respectively. Different core sizes are shown in solid ($D_{\rm core}=2.2$nm), dashed ($D_{\rm core}=3.0$nm), and dot-dashed ($D_{\rm core}=3.9$nm) lines.}
    \label{fig:crossover}
\end{figure*}

We also find that as the dephasing rate increases with increasing hybridization energy, $J$, deep in the adiabatic limit ($\gamma \gg 1$). This seems to be opposite to the behavior expected for the spin-boson model,\cite{thoss_self-consistent_2001} and can be traced to the inclusion of off-diagonal electron-phonon couplings terms in our model Hamiltonian ($V_{nm}^{\alpha}$). As shown in the Supporting Information Fig.~S7, the behavior of the population dynamics and the dephasing rates are consistent with the spin-boson model when the off-diagonal coupling terms are turned off ($V_{n\ne m}^\alpha=0$).  In addition, the presence of several donor and several acceptor states also affects the dephasing rates, particularly for $\gamma \gg 1$ where the $1P_e$-like donor/acceptor orbitals play a significant role in the charge transfer dynamics due to a smaller energy gap between $1S_e$ and $1P_e$-like states.

To further analyze the dephasing dynamics and delineate the modes that most strongly affect the electron transfer rates, we define the spectral density characterizing the electron-phonon interaction between states $\ket{\phi_n}$ and $\ket{\phi_m}$ ($n,m\in\mathcal{D},{A}$) as
\begin{equation}
    S_{nm}(\omega) = \pi\sum_\alpha\omega_\alpha\lambda^\alpha_{nm}\delta(\omega-\omega_\alpha)
\end{equation}
where $\lambda^\alpha_{nm} =  \frac{1}{2}\left(\frac{V_{nm}^\alpha}{\omega_\alpha}\right)^2$ is the reorganization energy for mode $\alpha$. In Fig.~\ref{fig:vibronic} panels (a) and (b) we plot the diagonal and off-diagonal spectral densities between the ground states of the donor and acceptor ($\ket{\phi_{D_1}}$ and $\ket{\phi_{A_1}}$) for the same systems shown in Fig.~\ref{fig:pop_panel}. The spectral densities are very structured with stronger coupling to the low-frequency acoustic modes ($\omega/2\pi<1.5$ THz) and weaker coupling to the high-frequency optical modes ($\omega/2\pi>4$ THz). We find that the overall magnitude of the diagonal spectral densities (panel (a)) decrease with increasing $\gamma$ while the off-diagonal spectral densities behave the opposite. Thus, setting the non-diagonal coupling terms to zero leads to a decrease in the dephasing rates with increasing  hybridization energies, which is not the case observed in Fig.~\ref{fig:pop_panel}, where the dephasing times are governed by the off-diagonal couplings.  In addition, we identify the modes that contribute the most to the dephasing dynamics (panels (c) and (d) of Fig.~\ref{fig:vibronic}). The lower frequency mode, $\omega_1$, involves a breathing motion while the higher frequency mode, $\omega_2$ involves a torsional motion. These vibrational modes are important to localize the charge and to facilitate the transfer dynamics.

In Fig.~\ref{fig:crossover} we provide a summary of the electron transfer rate constants across the nonadiabatic to adiabatic transition, and show how they depend on the neck and shell dimensions. In the adiabatic regime, we estimate the electron transfer rate constant by fitting the decay of the envelop of the donor population (see SI). Panel (a) of Fig.~\ref{fig:crossover} shows a scattered plot of the rate constants calculated for all the dimers considered in this work as a function of the ground state hybridization energy, $J$, at room temperature. As can be seen, the rate constants follows the $J^2$-Marcus theory dependence as $J\rightarrow 0$. For larger values of $J$, the population dynamics are no longer characterized by an over-damped exponential decay (unlike the Zusman limit), and the electron transfer is dominated by the dephasing times, which show a weaker dependence on $J$. The first order dependence of the dephasing rate on $J$ can be derived for the simple two-level spin-boson model, and is given by\cite{breuer_theory_2002} $k_{\rm dp}=2S(2J)\coth{(J/ k_{\rm B}T)}$, where $S(\omega)$ is the spectral density defined above. For the multi-state model considered in this work, the dephasing rate depends on other factors discussed previously, such as the off-diagonal spectral densities and contributions from higher-lying states, and increases with $J$ rather than decrease for the standard two-state model.  The solid black line in Fig.~\ref{fig:crossover}(a) is a fit to a connection formula for the total electron transfer rate, given by $k^{-1}=k_{\rm M}^{-1} + k_{\rm dp}^{-1}$\, with an empirical form for $k_{\rm dp}\propto\sqrt{J}$.

Fig.~\ref{fig:crossover} panels (b) and (c) show the dependence of $k$ on the neck width, $D_{\rm neck}$, and shell thickness, $D_{\rm shell}$, for different core diameters $D_{\rm core}$. We find that across the range of neck and shell dimensions that can be varied experimentally,\cite{cui_neck_2021} the electron transfer rate constants can be tuned over a broad range of values, from $\approx 10$ ps to $100$ fs.  This suggests that faster, room temperature, electron transfer devices require thinner shells,  wider necks, and larger cores. 

In conclusion, we outlined a theoretical framework to calculate charge transfer between two colloidal quantum dots connected by a bridge. Our approach used an atomistic model to describe the electronic structure, the normal modes of the donor/acceptor system, the electron-phonon couplings, and used a mixed quantum-classical mean-field Ehrenfest method to describe the time evolution of density matrix of the system.  We showed that by increasing the neck size, increasing the core size, and/or decreasing the shell thickness of the quantum dot building blocks, it is possible to control the hybridization energy and tune the system from the Marcus nonadiabatic, slow electron transfer regime to a coherent, adiabatic limit, even at elevated temperatures, with electron transfer times that are orders of magnitude faster and scale mildly with the hybridization energy. In all regimes, the charge transfer dynamics are mainly governed by the coupling of the excess electron to inter-dot and torsional acoustic modes. 

\begin{acknowledgments}
This work was supported by the NSF–BSF International Collaboration in the Division of Materials Research program, NSF grant number DMR-2026741 and BSF grant number 2020618. Methods used in this work were provided by the Center for Computational Study of Excited State Phenomena in Energy Materials (C2SEPEM), which is funded by the U.S. Department of Energy, Office of Science, Basic Energy Sciences, Materials Sciences and Engineering Division, via contract no. DE-AC02- 05CH11231, as part of the Computational Materials Sciences Program. Computational resources were provided in part by the National Energy Research Scientific Computing Center (NERSC), a U.S. Department of Energy Office of Science User Facility operated under contract no. DE-AC02- 05CH11231. D.J. acknowledges the support of the Computational Science Graduate Fellowship from the U.S. Department of Energy under grant no. DE-SC0019323.
\end{acknowledgments}

\section*{Reference}
\bibliographystyle{aipnum4-2}
\bibliography{main}

\begin{thebibliography}{30}%
\makeatletter
\providecommand \@ifxundefined [1]{%
 \@ifx{#1\undefined}
}%
\providecommand \@ifnum [1]{%
 \ifnum #1\expandafter \@firstoftwo
 \else \expandafter \@secondoftwo
 \fi
}%
\providecommand \@ifx [1]{%
 \ifx #1\expandafter \@firstoftwo
 \else \expandafter \@secondoftwo
 \fi
}%
\providecommand \natexlab [1]{#1}%
\providecommand \enquote  [1]{``#1''}%
\providecommand \bibnamefont  [1]{#1}%
\providecommand \bibfnamefont [1]{#1}%
\providecommand \citenamefont [1]{#1}%
\providecommand \href@noop [0]{\@secondoftwo}%
\providecommand \href [0]{\begingroup \@sanitize@url \@href}%
\providecommand \@href[1]{\@@startlink{#1}\@@href}%
\providecommand \@@href[1]{\endgroup#1\@@endlink}%
\providecommand \@sanitize@url [0]{\catcode `\\12\catcode `\$12\catcode
  `\&12\catcode `\#12\catcode `\^12\catcode `\_12\catcode `\%12\relax}%
\providecommand \@@startlink[1]{}%
\providecommand \@@endlink[0]{}%
\providecommand \url  [0]{\begingroup\@sanitize@url \@url }%
\providecommand \@url [1]{\endgroup\@href {#1}{\urlprefix }}%
\providecommand \urlprefix  [0]{URL }%
\providecommand \Eprint [0]{\href }%
\providecommand \doibase [0]{https://doi.org/}%
\providecommand \selectlanguage [0]{\@gobble}%
\providecommand \bibinfo  [0]{\@secondoftwo}%
\providecommand \bibfield  [0]{\@secondoftwo}%
\providecommand \translation [1]{[#1]}%
\providecommand \BibitemOpen [0]{}%
\providecommand \bibitemStop [0]{}%
\providecommand \bibitemNoStop [0]{.\EOS\space}%
\providecommand \EOS [0]{\spacefactor3000\relax}%
\providecommand \BibitemShut  [1]{\csname bibitem#1\endcsname}%
\let\auto@bib@innerbib\@empty
\bibitem [{\citenamefont {Marcus}(1956)}]{marcus_theory_1956}%
  \BibitemOpen
  \bibfield  {author} {\bibinfo {author} {\bibfnamefont {R.~A.}\ \bibnamefont
  {Marcus}},\ }\href {https://doi.org/10.1063/1.1742723} {\bibfield  {journal}
  {\bibinfo  {journal} {J. Chem. Phys.}\ }\textbf {\bibinfo {volume} {24}},\
  \bibinfo {pages} {966} (\bibinfo {year} {1956})}\BibitemShut {NoStop}%
\bibitem [{\citenamefont {Newton}\ and\ \citenamefont
  {Sutin}(1984)}]{newton_electron_1984}%
  \BibitemOpen
  \bibfield  {author} {\bibinfo {author} {\bibfnamefont {M.~D.}\ \bibnamefont
  {Newton}}\ and\ \bibinfo {author} {\bibfnamefont {N.}~\bibnamefont {Sutin}},\
  }\href {https://doi.org/10.1146/annurev.pc.35.100184.002253} {\bibfield
  {journal} {\bibinfo  {journal} {Annu. Rev. Phys. Chem.}\ }\textbf {\bibinfo
  {volume} {35}},\ \bibinfo {pages} {437} (\bibinfo {year} {1984})}\BibitemShut
  {NoStop}%
\bibitem [{\citenamefont {Jortner}(1976)}]{jortner_temperature_1976}%
  \BibitemOpen
  \bibfield  {author} {\bibinfo {author} {\bibfnamefont {J.}~\bibnamefont
  {Jortner}},\ }\href {https://doi.org/10.1063/1.432142} {\bibfield  {journal}
  {\bibinfo  {journal} {J. Chem. Phys.}\ }\textbf {\bibinfo {volume} {64}},\
  \bibinfo {pages} {4860} (\bibinfo {year} {1976})}\BibitemShut {NoStop}%
\bibitem [{\citenamefont {Redfield}(1965)}]{redfield_theory_1965}%
  \BibitemOpen
  \bibfield  {author} {\bibinfo {author} {\bibfnamefont {A.}~\bibnamefont
  {Redfield}},\ }in\ \href {https://doi.org/10.1016/B978-1-4832-3114-3.50007-6}
  {\emph {\bibinfo {booktitle} {Advances in {Magnetic} and {Optical}
  {Resonance}}}},\ Vol.~\bibinfo {volume} {1}\ (\bibinfo  {publisher}
  {Elsevier},\ \bibinfo {year} {1965})\ pp.\ \bibinfo {pages}
  {1--32}\BibitemShut {NoStop}%
\bibitem [{\citenamefont {Mohseni}\ \emph {et~al.}(2014)\citenamefont
  {Mohseni}, \citenamefont {Aspuru-Guzik}, \citenamefont {Rebentrost},
  \citenamefont {Shabani}, \citenamefont {Lloyd}, \citenamefont {Huelga},\ and\
  \citenamefont {Plenio}}]{engel_environment-assisted_2014}%
  \BibitemOpen
  \bibfield  {author} {\bibinfo {author} {\bibfnamefont {M.}~\bibnamefont
  {Mohseni}}, \bibinfo {author} {\bibfnamefont {A.}~\bibnamefont
  {Aspuru-Guzik}}, \bibinfo {author} {\bibfnamefont {P.}~\bibnamefont
  {Rebentrost}}, \bibinfo {author} {\bibfnamefont {A.}~\bibnamefont {Shabani}},
  \bibinfo {author} {\bibfnamefont {S.}~\bibnamefont {Lloyd}}, \bibinfo
  {author} {\bibfnamefont {S.~F.}\ \bibnamefont {Huelga}},\ and\ \bibinfo
  {author} {\bibfnamefont {M.~B.}\ \bibnamefont {Plenio}},\ }in\ \href
  {https://doi.org/10.1017/CBO9780511863189.009} {\emph {\bibinfo {booktitle}
  {Quantum {Effects} in {Biology}}}},\ \bibinfo {editor} {edited by\ \bibinfo
  {editor} {\bibfnamefont {G.~S.}\ \bibnamefont {Engel}}, \bibinfo {editor}
  {\bibfnamefont {M.~B.}\ \bibnamefont {Plenio}}, \bibinfo {editor}
  {\bibfnamefont {M.}~\bibnamefont {Mohseni}},\ and\ \bibinfo {editor}
  {\bibfnamefont {Y.}~\bibnamefont {Omar}}}\ (\bibinfo  {publisher} {Cambridge
  University Press},\ \bibinfo {address} {Cambridge},\ \bibinfo {year} {2014})\
  pp.\ \bibinfo {pages} {159--176}\BibitemShut {NoStop}%
\bibitem [{\citenamefont {Heinrich}\ \emph {et~al.}(2021)\citenamefont
  {Heinrich}, \citenamefont {Oliver}, \citenamefont {Vandersypen},
  \citenamefont {Ardavan}, \citenamefont {Sessoli}, \citenamefont {Loss},
  \citenamefont {Jayich}, \citenamefont {Fernandez-Rossier}, \citenamefont
  {Laucht},\ and\ \citenamefont {Morello}}]{heinrich_quantum-coherent_2021}%
  \BibitemOpen
  \bibfield  {author} {\bibinfo {author} {\bibfnamefont {A.~J.}\ \bibnamefont
  {Heinrich}}, \bibinfo {author} {\bibfnamefont {W.~D.}\ \bibnamefont
  {Oliver}}, \bibinfo {author} {\bibfnamefont {L.~M.~K.}\ \bibnamefont
  {Vandersypen}}, \bibinfo {author} {\bibfnamefont {A.}~\bibnamefont
  {Ardavan}}, \bibinfo {author} {\bibfnamefont {R.}~\bibnamefont {Sessoli}},
  \bibinfo {author} {\bibfnamefont {D.}~\bibnamefont {Loss}}, \bibinfo {author}
  {\bibfnamefont {A.~B.}\ \bibnamefont {Jayich}}, \bibinfo {author}
  {\bibfnamefont {J.}~\bibnamefont {Fernandez-Rossier}}, \bibinfo {author}
  {\bibfnamefont {A.}~\bibnamefont {Laucht}},\ and\ \bibinfo {author}
  {\bibfnamefont {A.}~\bibnamefont {Morello}},\ }\href
  {https://doi.org/10.1038/s41565-021-00994-1} {\bibfield  {journal} {\bibinfo
  {journal} {Nat. Nanotechnol.}\ }\textbf {\bibinfo {volume} {16}},\ \bibinfo
  {pages} {1318} (\bibinfo {year} {2021})}\BibitemShut {NoStop}%
\bibitem [{\citenamefont {Zusman}(1980)}]{zusman_outer-sphere_1980}%
  \BibitemOpen
  \bibfield  {author} {\bibinfo {author} {\bibfnamefont {L.}~\bibnamefont
  {Zusman}},\ }\href {https://doi.org/10.1016/0301-0104(80)85267-0} {\bibfield
  {journal} {\bibinfo  {journal} {Chem. Phys.}\ }\textbf {\bibinfo {volume}
  {49}},\ \bibinfo {pages} {295} (\bibinfo {year} {1980})}\BibitemShut
  {NoStop}%
\bibitem [{\citenamefont {Zusman}(1983)}]{zusman_theory_1983}%
  \BibitemOpen
  \bibfield  {author} {\bibinfo {author} {\bibfnamefont {L.}~\bibnamefont
  {Zusman}},\ }\href {https://doi.org/10.1016/0301-0104(83)85166-0} {\bibfield
  {journal} {\bibinfo  {journal} {Chem. Phys.}\ }\textbf {\bibinfo {volume}
  {80}},\ \bibinfo {pages} {29} (\bibinfo {year} {1983})}\BibitemShut {NoStop}%
\bibitem [{\citenamefont {Demadis}, \citenamefont {Hartshorn},\ and\
  \citenamefont {Meyer}(2001)}]{demadis_localized--delocalized_2001}%
  \BibitemOpen
  \bibfield  {author} {\bibinfo {author} {\bibfnamefont {K.~D.}\ \bibnamefont
  {Demadis}}, \bibinfo {author} {\bibfnamefont {C.~M.}\ \bibnamefont
  {Hartshorn}},\ and\ \bibinfo {author} {\bibfnamefont {T.~J.}\ \bibnamefont
  {Meyer}},\ }\href {https://doi.org/10.1021/cr990413m} {\bibfield  {journal}
  {\bibinfo  {journal} {Chem. Rev.}\ }\textbf {\bibinfo {volume} {101}},\
  \bibinfo {pages} {2655} (\bibinfo {year} {2001})}\BibitemShut {NoStop}%
\bibitem [{\citenamefont {Fiebig}\ \emph {et~al.}(2001)\citenamefont {Fiebig},
  \citenamefont {Stock}, \citenamefont {Lochbrunner},\ and\ \citenamefont
  {Riedle}}]{fiebig_femtosecond_2001}%
  \BibitemOpen
  \bibfield  {author} {\bibinfo {author} {\bibfnamefont {T.}~\bibnamefont
  {Fiebig}}, \bibinfo {author} {\bibfnamefont {K.}~\bibnamefont {Stock}},
  \bibinfo {author} {\bibfnamefont {S.}~\bibnamefont {Lochbrunner}},\ and\
  \bibinfo {author} {\bibfnamefont {E.}~\bibnamefont {Riedle}},\ }\href@noop {}
  {\bibfield  {journal} {\bibinfo  {journal} {Chem. Phys. Lett.}\ }\textbf
  {\bibinfo {volume} {345}},\ \bibinfo {pages} {8} (\bibinfo {year}
  {2001})}\BibitemShut {NoStop}%
\bibitem [{\citenamefont {Zhu}\ \emph {et~al.}(2021)\citenamefont {Zhu},
  \citenamefont {Qin}, \citenamefont {Meng}, \citenamefont {Mallick},
  \citenamefont {Gao}, \citenamefont {Chen}, \citenamefont {Cheng},
  \citenamefont {Tan}, \citenamefont {Xiao}, \citenamefont {Han}, \citenamefont
  {Sun},\ and\ \citenamefont {Liu}}]{zhu_crossover_2021}%
  \BibitemOpen
  \bibfield  {author} {\bibinfo {author} {\bibfnamefont {G.~Y.}\ \bibnamefont
  {Zhu}}, \bibinfo {author} {\bibfnamefont {Y.}~\bibnamefont {Qin}}, \bibinfo
  {author} {\bibfnamefont {M.}~\bibnamefont {Meng}}, \bibinfo {author}
  {\bibfnamefont {S.}~\bibnamefont {Mallick}}, \bibinfo {author} {\bibfnamefont
  {H.}~\bibnamefont {Gao}}, \bibinfo {author} {\bibfnamefont {X.}~\bibnamefont
  {Chen}}, \bibinfo {author} {\bibfnamefont {T.}~\bibnamefont {Cheng}},
  \bibinfo {author} {\bibfnamefont {Y.~N.}\ \bibnamefont {Tan}}, \bibinfo
  {author} {\bibfnamefont {X.}~\bibnamefont {Xiao}}, \bibinfo {author}
  {\bibfnamefont {M.~J.}\ \bibnamefont {Han}}, \bibinfo {author} {\bibfnamefont
  {M.~F.}\ \bibnamefont {Sun}},\ and\ \bibinfo {author} {\bibfnamefont {C.~Y.}\
  \bibnamefont {Liu}},\ }\href {https://doi.org/10.1038/s41467-020-20557-7}
  {\bibfield  {journal} {\bibinfo  {journal} {Nat. Commun.}\ }\textbf {\bibinfo
  {volume} {12}},\ \bibinfo {pages} {456} (\bibinfo {year} {2021})}\BibitemShut
  {NoStop}%
\bibitem [{\citenamefont {Cui}\ \emph {et~al.}(2019)\citenamefont {Cui},
  \citenamefont {Panfil}, \citenamefont {Koley}, \citenamefont {Shamalia},
  \citenamefont {Waiskopf}, \citenamefont {Remennik}, \citenamefont {Popov},
  \citenamefont {Oded},\ and\ \citenamefont {Banin}}]{cui_colloidal_2019}%
  \BibitemOpen
  \bibfield  {author} {\bibinfo {author} {\bibfnamefont {J.}~\bibnamefont
  {Cui}}, \bibinfo {author} {\bibfnamefont {Y.~E.}\ \bibnamefont {Panfil}},
  \bibinfo {author} {\bibfnamefont {S.}~\bibnamefont {Koley}}, \bibinfo
  {author} {\bibfnamefont {D.}~\bibnamefont {Shamalia}}, \bibinfo {author}
  {\bibfnamefont {N.}~\bibnamefont {Waiskopf}}, \bibinfo {author}
  {\bibfnamefont {S.}~\bibnamefont {Remennik}}, \bibinfo {author}
  {\bibfnamefont {I.}~\bibnamefont {Popov}}, \bibinfo {author} {\bibfnamefont
  {M.}~\bibnamefont {Oded}},\ and\ \bibinfo {author} {\bibfnamefont
  {U.}~\bibnamefont {Banin}},\ }\href
  {https://doi.org/10.1038/s41467-019-13349-1} {\bibfield  {journal} {\bibinfo
  {journal} {Nat. Commun.}\ }\textbf {\bibinfo {volume} {10}},\ \bibinfo
  {pages} {5401} (\bibinfo {year} {2019})}\BibitemShut {NoStop}%
\bibitem [{\citenamefont {Cui}\ \emph {et~al.}(2021)\citenamefont {Cui},
  \citenamefont {Koley}, \citenamefont {Panfil}, \citenamefont {Levi},
  \citenamefont {Ossia}, \citenamefont {Waiskopf}, \citenamefont {Remennik},
  \citenamefont {Oded},\ and\ \citenamefont {Banin}}]{cui_neck_2021}%
  \BibitemOpen
  \bibfield  {author} {\bibinfo {author} {\bibfnamefont {J.}~\bibnamefont
  {Cui}}, \bibinfo {author} {\bibfnamefont {S.}~\bibnamefont {Koley}}, \bibinfo
  {author} {\bibfnamefont {Y.~E.}\ \bibnamefont {Panfil}}, \bibinfo {author}
  {\bibfnamefont {A.}~\bibnamefont {Levi}}, \bibinfo {author} {\bibfnamefont
  {Y.}~\bibnamefont {Ossia}}, \bibinfo {author} {\bibfnamefont
  {N.}~\bibnamefont {Waiskopf}}, \bibinfo {author} {\bibfnamefont
  {S.}~\bibnamefont {Remennik}}, \bibinfo {author} {\bibfnamefont
  {M.}~\bibnamefont {Oded}},\ and\ \bibinfo {author} {\bibfnamefont
  {U.}~\bibnamefont {Banin}},\ }\href {https://doi.org/10.1021/jacs.1c08863}
  {\bibfield  {journal} {\bibinfo  {journal} {J. Am. Chem. Soc.}\ }\textbf
  {\bibinfo {volume} {143}},\ \bibinfo {pages} {19816} (\bibinfo {year}
  {2021})}\BibitemShut {NoStop}%
\bibitem [{\citenamefont {Jasrasaria}\ and\ \citenamefont
  {Rabani}(2021)}]{jasrasaria_interplay_2021}%
  \BibitemOpen
  \bibfield  {author} {\bibinfo {author} {\bibfnamefont {D.}~\bibnamefont
  {Jasrasaria}}\ and\ \bibinfo {author} {\bibfnamefont {E.}~\bibnamefont
  {Rabani}},\ }\href {https://doi.org/10.1021/acs.nanolett.1c02953} {\bibfield
  {journal} {\bibinfo  {journal} {Nano Lett.}\ }\textbf {\bibinfo {volume}
  {21}},\ \bibinfo {pages} {8741} (\bibinfo {year} {2021})}\BibitemShut
  {NoStop}%
\bibitem [{\citenamefont {Wang}\ and\ \citenamefont
  {Zunger}(1994)}]{wang_electronic_1994}%
  \BibitemOpen
  \bibfield  {author} {\bibinfo {author} {\bibfnamefont {L.~W.}\ \bibnamefont
  {Wang}}\ and\ \bibinfo {author} {\bibfnamefont {A.}~\bibnamefont {Zunger}},\
  }\href {https://doi.org/10.1021/j100059a032} {\bibfield  {journal} {\bibinfo
  {journal} {J. Phys. Chem.}\ }\textbf {\bibinfo {volume} {98}},\ \bibinfo
  {pages} {2158} (\bibinfo {year} {1994})}\BibitemShut {NoStop}%
\bibitem [{\citenamefont {Rabani}\ \emph {et~al.}(1999)\citenamefont {Rabani},
  \citenamefont {Hetényi}, \citenamefont {Berne},\ and\ \citenamefont
  {Brus}}]{rabani_electronic_1999}%
  \BibitemOpen
  \bibfield  {author} {\bibinfo {author} {\bibfnamefont {E.}~\bibnamefont
  {Rabani}}, \bibinfo {author} {\bibfnamefont {B.}~\bibnamefont {Hetényi}},
  \bibinfo {author} {\bibfnamefont {B.~J.}\ \bibnamefont {Berne}},\ and\
  \bibinfo {author} {\bibfnamefont {L.~E.}\ \bibnamefont {Brus}},\ }\href
  {https://doi.org/10.1063/1.478431} {\bibfield  {journal} {\bibinfo  {journal}
  {J. Chem. Phys.}\ }\textbf {\bibinfo {volume} {110}},\ \bibinfo {pages}
  {5355} (\bibinfo {year} {1999})}\BibitemShut {NoStop}%
\bibitem [{\citenamefont {Wall}\ and\ \citenamefont
  {Neuhauser}(1995)}]{wall_extraction_1995}%
  \BibitemOpen
  \bibfield  {author} {\bibinfo {author} {\bibfnamefont {M.~R.}\ \bibnamefont
  {Wall}}\ and\ \bibinfo {author} {\bibfnamefont {D.}~\bibnamefont
  {Neuhauser}},\ }\href {https://doi.org/10.1063/1.468999} {\bibfield
  {journal} {\bibinfo  {journal} {J. Chem. Phys.}\ }\textbf {\bibinfo {volume}
  {102}},\ \bibinfo {pages} {8011} (\bibinfo {year} {1995})}\BibitemShut
  {NoStop}%
\bibitem [{\citenamefont {Toledo}\ and\ \citenamefont
  {Rabani}(2002)}]{toledo_very_2002}%
  \BibitemOpen
  \bibfield  {author} {\bibinfo {author} {\bibfnamefont {S.}~\bibnamefont
  {Toledo}}\ and\ \bibinfo {author} {\bibfnamefont {E.}~\bibnamefont
  {Rabani}},\ }\href {https://doi.org/10.1006/jcph.2002.7090} {\bibfield
  {journal} {\bibinfo  {journal} {J. Comput. Phys.}\ }\textbf {\bibinfo
  {volume} {180}},\ \bibinfo {pages} {256} (\bibinfo {year}
  {2002})}\BibitemShut {NoStop}%
\bibitem [{\citenamefont {Foster}\ and\ \citenamefont
  {Boys}(1960)}]{foster_canonical_1960}%
  \BibitemOpen
  \bibfield  {author} {\bibinfo {author} {\bibfnamefont {J.~M.}\ \bibnamefont
  {Foster}}\ and\ \bibinfo {author} {\bibfnamefont {S.~F.}\ \bibnamefont
  {Boys}},\ }\href {https://doi.org/10.1103/RevModPhys.32.300} {\bibfield
  {journal} {\bibinfo  {journal} {Rev. Mod. Phys.}\ }\textbf {\bibinfo {volume}
  {32}},\ \bibinfo {pages} {300} (\bibinfo {year} {1960})}\BibitemShut
  {NoStop}%
\bibitem [{\citenamefont {Kleier}\ \emph {et~al.}(1974)\citenamefont {Kleier},
  \citenamefont {Halgren}, \citenamefont {Hall},\ and\ \citenamefont
  {Lipscomb}}]{kleier_localized_1974}%
  \BibitemOpen
  \bibfield  {author} {\bibinfo {author} {\bibfnamefont {D.~A.}\ \bibnamefont
  {Kleier}}, \bibinfo {author} {\bibfnamefont {T.~A.}\ \bibnamefont {Halgren}},
  \bibinfo {author} {\bibfnamefont {J.~H.}\ \bibnamefont {Hall}},\ and\
  \bibinfo {author} {\bibfnamefont {W.~N.}\ \bibnamefont {Lipscomb}},\ }\href
  {https://doi.org/10.1063/1.1681683} {\bibfield  {journal} {\bibinfo
  {journal} {J. Chem. Phys.}\ }\textbf {\bibinfo {volume} {61}},\ \bibinfo
  {pages} {3905} (\bibinfo {year} {1974})}\BibitemShut {NoStop}%
\bibitem [{\citenamefont {Zhou}\ \emph {et~al.}(2013)\citenamefont {Zhou},
  \citenamefont {Ward}, \citenamefont {Martin}, \citenamefont {van Swol},
  \citenamefont {Cruz-Campa},\ and\ \citenamefont
  {Zubia}}]{zhou_stillinger-weber_2013}%
  \BibitemOpen
  \bibfield  {author} {\bibinfo {author} {\bibfnamefont {X.~W.}\ \bibnamefont
  {Zhou}}, \bibinfo {author} {\bibfnamefont {D.~K.}\ \bibnamefont {Ward}},
  \bibinfo {author} {\bibfnamefont {J.~E.}\ \bibnamefont {Martin}}, \bibinfo
  {author} {\bibfnamefont {F.~B.}\ \bibnamefont {van Swol}}, \bibinfo {author}
  {\bibfnamefont {J.~L.}\ \bibnamefont {Cruz-Campa}},\ and\ \bibinfo {author}
  {\bibfnamefont {D.}~\bibnamefont {Zubia}},\ }\href
  {https://doi.org/10.1103/PhysRevB.88.085309} {\bibfield  {journal} {\bibinfo
  {journal} {Phys. Rev. B}\ }\textbf {\bibinfo {volume} {88}},\ \bibinfo
  {pages} {085309} (\bibinfo {year} {2013})}\BibitemShut {NoStop}%
\bibitem [{\citenamefont {Crespo-Otero}\ and\ \citenamefont
  {Barbatti}(2018)}]{crespo-otero_recent_2018}%
  \BibitemOpen
  \bibfield  {author} {\bibinfo {author} {\bibfnamefont {R.}~\bibnamefont
  {Crespo-Otero}}\ and\ \bibinfo {author} {\bibfnamefont {M.}~\bibnamefont
  {Barbatti}},\ }\href {https://doi.org/10.1021/acs.chemrev.7b00577} {\bibfield
   {journal} {\bibinfo  {journal} {Chemical Reviews}\ }\textbf {\bibinfo
  {volume} {118}},\ \bibinfo {pages} {7026} (\bibinfo {year}
  {2018})}\BibitemShut {NoStop}%
\bibitem [{\citenamefont {McLachlan}(1964)}]{mclachlan_variational_1964}%
  \BibitemOpen
  \bibfield  {author} {\bibinfo {author} {\bibfnamefont {A.}~\bibnamefont
  {McLachlan}},\ }\href {https://doi.org/10.1080/00268976400100041} {\bibfield
  {journal} {\bibinfo  {journal} {Mol. Phys.}\ }\textbf {\bibinfo {volume}
  {8}},\ \bibinfo {pages} {39} (\bibinfo {year} {1964})}\BibitemShut {NoStop}%
\bibitem [{\citenamefont {Nijjar}, \citenamefont {Jankowska},\ and\
  \citenamefont {Prezhdo}(2019)}]{nijjar_ehrenfest_2019}%
  \BibitemOpen
  \bibfield  {author} {\bibinfo {author} {\bibfnamefont {P.}~\bibnamefont
  {Nijjar}}, \bibinfo {author} {\bibfnamefont {J.}~\bibnamefont {Jankowska}},\
  and\ \bibinfo {author} {\bibfnamefont {O.~V.}\ \bibnamefont {Prezhdo}},\
  }\href {https://doi.org/10.1063/1.5095810} {\bibfield  {journal} {\bibinfo
  {journal} {J. Chem. Phys.}\ }\textbf {\bibinfo {volume} {150}},\ \bibinfo
  {pages} {204124} (\bibinfo {year} {2019})}\BibitemShut {NoStop}%
\bibitem [{\citenamefont {Egorov}, \citenamefont {Rabani},\ and\ \citenamefont
  {Berne}(1998)}]{egorov_vibronic_1998}%
  \BibitemOpen
  \bibfield  {author} {\bibinfo {author} {\bibfnamefont {S.~A.}\ \bibnamefont
  {Egorov}}, \bibinfo {author} {\bibfnamefont {E.}~\bibnamefont {Rabani}},\
  and\ \bibinfo {author} {\bibfnamefont {B.~J.}\ \bibnamefont {Berne}},\ }\href
  {https://doi.org/10.1063/1.475512} {\bibfield  {journal} {\bibinfo  {journal}
  {J. Chem. Phys.}\ }\textbf {\bibinfo {volume} {108}},\ \bibinfo {pages}
  {1407} (\bibinfo {year} {1998})}\BibitemShut {NoStop}%
\bibitem [{\citenamefont {Egorov}, \citenamefont {Rabani},\ and\ \citenamefont
  {Berne}(1999)}]{egorov_nonradiative_1999}%
  \BibitemOpen
  \bibfield  {author} {\bibinfo {author} {\bibfnamefont {S.~A.}\ \bibnamefont
  {Egorov}}, \bibinfo {author} {\bibfnamefont {E.}~\bibnamefont {Rabani}},\
  and\ \bibinfo {author} {\bibfnamefont {B.~J.}\ \bibnamefont {Berne}},\ }\href
  {https://doi.org/10.1063/1.478420} {\bibfield  {journal} {\bibinfo  {journal}
  {J. Chem. Phys.}\ }\textbf {\bibinfo {volume} {110}},\ \bibinfo {pages}
  {5238} (\bibinfo {year} {1999})}\BibitemShut {NoStop}%
\bibitem [{\citenamefont {Shemetulskis}\ and\ \citenamefont
  {Loring}(1992)}]{shemetulskis_semiclassical_1992}%
  \BibitemOpen
  \bibfield  {author} {\bibinfo {author} {\bibfnamefont {N.~E.}\ \bibnamefont
  {Shemetulskis}}\ and\ \bibinfo {author} {\bibfnamefont {R.~F.}\ \bibnamefont
  {Loring}},\ }\href {https://doi.org/10.1063/1.463248} {\bibfield  {journal}
  {\bibinfo  {journal} {J. Chem. Phys.}\ }\textbf {\bibinfo {volume} {97}},\
  \bibinfo {pages} {1217} (\bibinfo {year} {1992})}\BibitemShut {NoStop}%
\bibitem [{Note1()}]{Note1}%
  \BibitemOpen
  \bibinfo {note} {Quantum mechanical test calculations using the
  multiconfiguration time-dependent Hartree (MCTDH) method show good agreement
  with the results of the Ehrenfest method.}\BibitemShut {Stop}%
\bibitem [{\citenamefont {Thoss}, \citenamefont {Wang},\ and\ \citenamefont
  {Miller}(2001)}]{thoss_self-consistent_2001}%
  \BibitemOpen
  \bibfield  {author} {\bibinfo {author} {\bibfnamefont {M.}~\bibnamefont
  {Thoss}}, \bibinfo {author} {\bibfnamefont {H.}~\bibnamefont {Wang}},\ and\
  \bibinfo {author} {\bibfnamefont {W.~H.}\ \bibnamefont {Miller}},\ }\href
  {https://doi.org/10.1063/1.1385562} {\bibfield  {journal} {\bibinfo
  {journal} {J. Chem. Phys.}\ }\textbf {\bibinfo {volume} {115}},\ \bibinfo
  {pages} {2991} (\bibinfo {year} {2001})}\BibitemShut {NoStop}%
\bibitem [{\citenamefont {Breuer}\ and\ \citenamefont
  {Petruccione}(2002)}]{breuer_theory_2002}%
  \BibitemOpen
  \bibfield  {author} {\bibinfo {author} {\bibfnamefont {H.-P.}\ \bibnamefont
  {Breuer}}\ and\ \bibinfo {author} {\bibfnamefont {F.}~\bibnamefont
  {Petruccione}},\ }\href@noop {} {\emph {\bibinfo {title} {The {Theory} of
  {Open} {Quantum} {Systems}}}}\ (\bibinfo  {publisher} {Oxford University
  Press},\ \bibinfo {year} {2002})\BibitemShut {NoStop}%
\end{thebibliography}%


\providecommand{\latin}[1]{#1}
\makeatletter
\providecommand{\doi}
  {\begingroup\let\do\@makeother\dospecials
  \catcode`\{=1 \catcode`\}=2 \doi@aux}
\providecommand{\doi@aux}[1]{\endgroup\texttt{#1}}
\makeatother
\providecommand*\mcitethebibliography{\thebibliography}
\csname @ifundefined\endcsname{endmcitethebibliography}
  {\let\endmcitethebibliography\endthebibliography}{}
\begin{mcitethebibliography}{12}
\providecommand*\natexlab[1]{#1}
\providecommand*\mciteSetBstSublistMode[1]{}
\providecommand*\mciteSetBstMaxWidthForm[2]{}
\providecommand*\mciteBstWouldAddEndPuncttrue
  {\def\EndOfBibitem{\unskip.}}
\providecommand*\mciteBstWouldAddEndPunctfalse
  {\let\EndOfBibitem\relax}
\providecommand*\mciteSetBstMidEndSepPunct[3]{}
\providecommand*\mciteSetBstSublistLabelBeginEnd[3]{}
\providecommand*\EndOfBibitem{}
\mciteSetBstSublistMode{f}
\mciteSetBstMaxWidthForm{subitem}{(\alph{mcitesubitemcount})}
\mciteSetBstSublistLabelBeginEnd
  {\mcitemaxwidthsubitemform\space}
  {\relax}
  {\relax}

\bibitem[Zhou \latin{et~al.}(2013)Zhou, Ward, Martin, van Swol, Cruz-Campa, and
  Zubia]{zhou_stillinger-weber_2013}
Zhou,~X.~W.; Ward,~D.~K.; Martin,~J.~E.; van Swol,~F.~B.; Cruz-Campa,~J.~L.;
  Zubia,~D. Stillinger-{Weber} potential for the {II}-{VI} elements
  {Zn}-{Cd}-{Hg}-{S}-{Se}-{Te}. \emph{Phys. Rev. B} \textbf{2013}, \emph{88},
  085309\relax
\mciteBstWouldAddEndPuncttrue
\mciteSetBstMidEndSepPunct{\mcitedefaultmidpunct}
{\mcitedefaultendpunct}{\mcitedefaultseppunct}\relax
\EndOfBibitem
\bibitem[Wang and Zunger(1994)Wang, and Zunger]{wang_electronic_1994}
Wang,~L.~W.; Zunger,~A. Electronic {Structure} {Pseudopotential} {Calculations}
  of {Large} (.apprx.1000 {Atoms}) {Si} {Quantum} {Dots}. \emph{J. Phys. Chem.}
  \textbf{1994}, \emph{98}, 2158--2165\relax
\mciteBstWouldAddEndPuncttrue
\mciteSetBstMidEndSepPunct{\mcitedefaultmidpunct}
{\mcitedefaultendpunct}{\mcitedefaultseppunct}\relax
\EndOfBibitem
\bibitem[Rabani \latin{et~al.}(1999)Rabani, Hetényi, Berne, and
  Brus]{rabani_electronic_1999}
Rabani,~E.; Hetényi,~B.; Berne,~B.~J.; Brus,~L.~E. Electronic properties of
  {CdSe} nanocrystals in the absence and presence of a dielectric medium.
  \emph{J. Chem. Phys.} \textbf{1999}, \emph{110}, 5355--5369\relax
\mciteBstWouldAddEndPuncttrue
\mciteSetBstMidEndSepPunct{\mcitedefaultmidpunct}
{\mcitedefaultendpunct}{\mcitedefaultseppunct}\relax
\EndOfBibitem
\bibitem[Wall and Neuhauser(1995)Wall, and Neuhauser]{wall_extraction_1995}
Wall,~M.~R.; Neuhauser,~D. Extraction, through filter‐diagonalization, of
  general quantum eigenvalues or classical normal mode frequencies from a small
  number of residues or a short‐time segment of a signal. {I}. {Theory} and
  application to a quantum‐dynamics model. \emph{J. Chem. Phys.}
  \textbf{1995}, \emph{102}, 8011--8022\relax
\mciteBstWouldAddEndPuncttrue
\mciteSetBstMidEndSepPunct{\mcitedefaultmidpunct}
{\mcitedefaultendpunct}{\mcitedefaultseppunct}\relax
\EndOfBibitem
\bibitem[Toledo and Rabani(2002)Toledo, and Rabani]{toledo_very_2002}
Toledo,~S.; Rabani,~E. Very {Large} {Electronic} {Structure} {Calculations}
  {Using} an {Out}-of-{Core} {Filter}-{Diagonalization} {Method}. \emph{J.
  Comput. Phys.} \textbf{2002}, \emph{180}, 256--269\relax
\mciteBstWouldAddEndPuncttrue
\mciteSetBstMidEndSepPunct{\mcitedefaultmidpunct}
{\mcitedefaultendpunct}{\mcitedefaultseppunct}\relax
\EndOfBibitem
\bibitem[Wang \latin{et~al.}(1999)Wang, Kim, and Zunger]{wang_electronic_1999}
Wang,~L.-W.; Kim,~J.; Zunger,~A. Electronic structures of [110]-faceted
  self-assembled pyramidal {InAs}/{GaAs} quantum dots. \emph{Phys. Rev. B}
  \textbf{1999}, \emph{59}, 5678--5687\relax
\mciteBstWouldAddEndPuncttrue
\mciteSetBstMidEndSepPunct{\mcitedefaultmidpunct}
{\mcitedefaultendpunct}{\mcitedefaultseppunct}\relax
\EndOfBibitem
\bibitem[Foster and Boys(1960)Foster, and Boys]{foster_canonical_1960}
Foster,~J.~M.; Boys,~S.~F. Canonical {Configurational} {Interaction}
  {Procedure}. \emph{Rev. Mod. Phys.} \textbf{1960}, \emph{32}, 300--302\relax
\mciteBstWouldAddEndPuncttrue
\mciteSetBstMidEndSepPunct{\mcitedefaultmidpunct}
{\mcitedefaultendpunct}{\mcitedefaultseppunct}\relax
\EndOfBibitem
\bibitem[Kleier \latin{et~al.}(1974)Kleier, Halgren, Hall, and
  Lipscomb]{kleier_localized_1974}
Kleier,~D.~A.; Halgren,~T.~A.; Hall,~J.~H.; Lipscomb,~W.~N. Localized molecular
  orbitals for polyatomic molecules. {I}. {A} comparison of the
  {Edmiston}‐{Ruedenberg} and {Boys} localization methods. \emph{J. Chem.
  Phys.} \textbf{1974}, \emph{61}, 3905--3919\relax
\mciteBstWouldAddEndPuncttrue
\mciteSetBstMidEndSepPunct{\mcitedefaultmidpunct}
{\mcitedefaultendpunct}{\mcitedefaultseppunct}\relax
\EndOfBibitem
\bibitem[Jasrasaria and Rabani(2021)Jasrasaria, and
  Rabani]{jasrasaria_interplay_2021}
Jasrasaria,~D.; Rabani,~E. Interplay of {Surface} and {Interior} {Modes} in
  {Exciton}–{Phonon} {Coupling} at the {Nanoscale}. \emph{Nano Lett.}
  \textbf{2021}, \emph{21}, 8741--8748\relax
\mciteBstWouldAddEndPuncttrue
\mciteSetBstMidEndSepPunct{\mcitedefaultmidpunct}
{\mcitedefaultendpunct}{\mcitedefaultseppunct}\relax
\EndOfBibitem
\bibitem[Berkelbach \latin{et~al.}(2012)Berkelbach, Reichman, and
  Markland]{berkelbach_reduced_2012}
Berkelbach,~T.~C.; Reichman,~D.~R.; Markland,~T.~E. Reduced density matrix
  hybrid approach: {An} efficient and accurate method for adiabatic and
  non-adiabatic quantum dynamics. \emph{J. Chem. Phys.} \textbf{2012},
  \emph{136}, 034113\relax
\mciteBstWouldAddEndPuncttrue
\mciteSetBstMidEndSepPunct{\mcitedefaultmidpunct}
{\mcitedefaultendpunct}{\mcitedefaultseppunct}\relax
\EndOfBibitem
\bibitem[Kapral(2006)]{kapral_progress_2006}
Kapral,~R. {Progress} {in} {the} {theory} {of} {mixed} {quantum}-{classical}
  {dynamics}. \emph{Annu. Rev. Phys. Chem.} \textbf{2006}, \emph{57},
  129--157\relax
\mciteBstWouldAddEndPuncttrue
\mciteSetBstMidEndSepPunct{\mcitedefaultmidpunct}
{\mcitedefaultendpunct}{\mcitedefaultseppunct}\relax
\EndOfBibitem
\end{mcitethebibliography}

\end{document}









\section{Nanostructure configurations}
The core-shell colloidal quantum dot (CQD) nanocrystals (NCs) are constructed by adding CdS shells to a CdSe core which is cleaved from a large crystal with a lattice constant of bulk wurtzite CdSe ($a=4.3$\AA, $c=\sqrt{\frac{8}{3}}a$). The core diameters and shell thickness for single NC are denoted as $D_{\rm core}$ and $D_{\rm shell}$ respectively. The CQD dimers are constructed by attaching two NCs either through the $[100]$ symmetric orientation or $[001]$ asymmetric orientation. The neck/bridge of the dimer can then be widened by adding additional CdS layers the connection area between two NCs, and the neck width is denoted as $D_{\rm neck}$. Tables. \ref{tab:1} and \ref{tab:2} summarize the different combinations of $D_{\rm neck}$, $D_{\rm shell}$ and $D_{\rm core}$ in two orientations used in this study. The structures were minimized with Stillinger-Weber force field parameterized for II-VI nanostructures\cite{zhou_stillinger-weber_2013} using the conjugate gradient minimization implemented in LAMMPS.

\begin{table}[ht]
\begin{tabular}{|c|c|c|c|c|c|c|}
\hline
\caption{Dimensions of dimers (in nm) by controlling $D_{\rm core}$ and $D_{\rm neck}$. The total dimension $D=D_{\rm core}+2D_{\rm shell}$. }
\label{tab:1}
Orientations               & Label & $D_{\rm core}$ & $D_{\rm shell}$ &  $D$ & $D_{\rm neck}$ & Formula          \\ \hline
\multirow{9}{*}{{[}100{]}} & (1a)  & 2.2  & 1.7   & 5.7 & 2.4  & $\text{Cd}_{5330}\text{Se}_{504}\text{S}_{4826}$ \\ \cline{2-7} 
                           & (1b)  & 3.0  & 1.3   & 5.7 & 2.4  & $\text{Cd}_{5330}\text{Se}_{966}\text{S}_{4364}$ \\ \cline{2-7} 
                           & (1c)  & 3.9  & 0.9   & 5.7 & 2.4  & $\text{Cd}_{5330}\text{Se}_{216}\text{S}_{5116}$ \\ \cline{2-7} 
                           & (1d)  & 2.2  & 1.7   & 5.7 & 3.1  & $\text{Cd}_{5510}\text{Se}_{216}\text{S}_{5294}$ \\ \cline{2-7} 
                           & (1e)  & 3.0  & 1.3   & 5.7 & 3.1  & $\text{Cd}_{5510}\text{Se}_{504}\text{S}_{5006}$ \\ \cline{2-7} 
                           & (1f)  & 3.9  & 0.9   & 5.7 & 3.1  & $\text{Cd}_{5510}\text{Se}_{966}\text{S}_{4544}$ \\ \cline{2-7} 
                           & (1g)  & 2.2  & 1.7   & 5.7 & 3.8  & $\text{Cd}_{5764}\text{Se}_{216}\text{S}_{5548}$ \\ \cline{2-7} 
                           & (1h)  & 3.0  & 1.3   & 5.7 & 3.8  & $\text{Cd}_{5764}\text{Se}_{504}\text{S}_{5260}$ \\ \cline{2-7} 
                           & (1i)  & 3.9  & 0.9   & 5.7 & 3.8  & $\text{Cd}_{5764}\text{Se}_{966}\text{S}_{4798}$ \\ \hline
\multirow{9}{*}{{[}001{]}} & (2a)  & 2.2  & 1.7   & 5.7 & 2.8  & $\text{Cd}_{5202}\text{Se}_{216}\text{S}_{4986}$ \\ \cline{2-7} 
                           & (2b)  & 3.0  & 1.3   & 5.7 & 2.8  & $\text{Cd}_{5202}\text{Se}_{504}\text{S}_{4698}$ \\ \cline{2-7} 
                           & (2c)  & 3.9  & 0.9   & 5.7 & 2.8  & $\text{Cd}_{5202}\text{Se}_{966}\text{S}_{4236}$ \\ \cline{2-7} 
                           & (2d)  & 2.2  & 1.7   & 5.7 & 3.5  & $\text{Cd}_{5283}\text{Se}_{216}\text{S}_{5067}$ \\ \cline{2-7} 
                           & (2e)  & 3.0  & 1.3   & 5.7 & 3.5  & $\text{Cd}_{5283}\text{Se}_{504}\text{S}_{4779}$ \\ \cline{2-7} 
                           & (2f)  & 3.9  & 0.9   & 5.7 & 3.5  & $\text{Cd}_{5283}\text{Se}_{966}\text{S}_{4317}$ \\ \cline{2-7} 
                           & (2g)  & 2.2  & 1.7   & 5.7 & 4.2  & $\text{Cd}_{5639}\text{Se}_{216}\text{S}_{5423}$ \\ \cline{2-7} 
                           & (2h)  & 3.0  & 1.3   & 5.7 & 4.2  & $\text{Cd}_{5639}\text{Se}_{504}\text{S}_{5135}$ \\ \cline{2-7} 
                           & (2i)  & 3.9  & 0.9   & 5.7 & 4.2  & $\text{Cd}_{5639}\text{Se}_{966}\text{S}_{4673}$ \\ \hline
\end{tabular}
\end{table}

\begin{table}[ht]
\begin{tabular}{|c|c|c|c|c|c|c|}
\hline
\caption{Dimensions of dimers (in nm) by controlling $D_{\rm core}$ and $D_{\rm shell}$. }
\label{tab:2}
Orientations               & Label & $D_{\rm core}$ & $D_{\rm shell}$ &  $D$ & $D_{\rm neck}$ & Formula          \\ \hline
\multirow{9}{*}{{[}100{]}} & (3a)  & 2.2  & 1.4   & 5.0 & 2.5  & $\text{Cd}_{3812}\text{Se}_{216}\text{S}_{3569}$ \\ \cline{2-7} 
                           & (3b)  & 3.0  & 1.3   & 5.7 & 2.9  & $\text{Cd}_{5330}\text{Se}_{504}\text{S}_{4826}$ \\ \cline{2-7} 
                           & (3c)  & 3.9  & 1.2   & 6.3 & 3.3  & $\text{Cd}_{7226}\text{Se}_{966}\text{S}_{6260}$ \\ \cline{2-7} 
                           & (3d)  & 2.2  & 1.1   & 4.3 & 2.1  & $\text{Cd}_{2630}\text{Se}_{216}\text{S}_{2414}$ \\ \cline{2-7} 
                           & (3e)  & 3.0  & 0.9   & 4.7 & 2.5  & $\text{Cd}_{3660}\text{Se}_{504}\text{S}_{3156}$ \\ \cline{2-7} 
                           & (3f)  & 3.9  & 0.9   & 5.7 & 2.9  & $\text{Cd}_{5330}\text{Se}_{966}\text{S}_{4364}$ \\ \cline{2-7} 
                           & (3g)  & 2.2  & 0.7   & 3.6 & 1.4  & $\text{Cd}_{1742}\text{Se}_{216}\text{S}_{1523}$ \\ \cline{2-7} 
                           & (3h)  & 3.0  & 0.7   & 4.3 & 2.1  & $\text{Cd}_{2478}\text{Se}_{504}\text{S}_{1974}$ \\ \cline{2-7} 
                           & (3i)  & 3.9  & 0.6   & 5.0 & 2.5  & $\text{Cd}_{3812}\text{Se}_{966}\text{S}_{2846}$ \\ \hline
\multirow{9}{*}{{[}001{]}} & (4a)  & 2.2  & 1.4   & 5.0 & 2.5  & $\text{Cd}_{3660}\text{Se}_{216}\text{S}_{3444}$ \\ \cline{2-7} 
                           & (4b)  & 3.0  & 1.3   & 5.7 & 2.9  & $\text{Cd}_{5202}\text{Se}_{504}\text{S}_{4698}$ \\ \cline{2-7} 
                           & (4c)  & 3.9  & 1.2   & 6.3 & 3.3  & $\text{Cd}_{7125}\text{Se}_{966}\text{S}_{6159}$ \\ \cline{2-7} 
                           & (4d)  & 2.2  & 1.1   & 4.3 & 2.1  & $\text{Cd}_{2457}\text{Se}_{216}\text{S}_{2241}$ \\ \cline{2-7} 
                           & (4e)  & 3.0  & 0.9   & 4.7 & 2.5  & $\text{Cd}_{3660}\text{Se}_{504}\text{S}_{3156}$ \\ \cline{2-7} 
                           & (4f)  & 3.9  & 0.9   & 5.7 & 2.9  & $\text{Cd}_{5202}\text{Se}_{966}\text{S}_{4236}$ \\ \cline{2-7} 
                           & (4g)  & 2.2  & 0.7   & 3.6 & 1.4  & $\text{Cd}_{1545}\text{Se}_{216}\text{S}_{1329}$ \\ \cline{2-7} 
                           & (4h)  & 3.0  & 0.7   & 4.3 & 2.1  & $\text{Cd}_{2457}\text{Se}_{504}\text{S}_{1953}$ \\ \cline{2-7} 
                           & (4i)  & 3.9  & 0.6   & 5.0 & 2.5  & $\text{Cd}_{3660}\text{Se}_{966}\text{S}_{2694}$ \\ \hline
\end{tabular}
\end{table}

\section{Quasi-electron states calculations and localization}
The semi-empirical pseudopotential model~\cite{wang_electronic_1994,rabani_electronic_1999} was used to describe the quasi-electron Hamiltonian and the filter diagonalization technique~\cite{wall_extraction_1995,toledo_very_2002} was applied to calculate the eigenstates of the dimer near the bottom of the conduction band. The local screened strain-dependent pseudopotentials following Wang \textit{et al.}\cite{wang_electronic_1999} have the functional form in the momentum space
\begin{equation}
\Tilde{\nu}(q)=a_0\left[1+a_4 \operatorname{Tr} \epsilon\right] \frac{q^2-a_1}{a_2 \exp \left(a_3 q^2\right)-1}
\end{equation}
where $\epsilon$ is the strain tensor, and the parameters for Cd, Se, and S are collected in Table \ref{tab:3}. All parameters were fitted to reproduce bulk band structures, band gaps, and effective masses. The real-space quasi-electron Hamiltonian $\hat{h}_{\rm QP}(\mathbf{r})$ is given by 
\begin{equation}
    \hat{h}_{\rm QP}(\mathbf{r}) = -\frac{1}{2}\nabla^2_{\mathbf{r}} + \sum_\mu \nu_\mu\left(|\mathbf{r}-\mathbf{R}_{0,\mu}|\right),
\end{equation}
where $\nu_\mu$ is the real-space pseudopotential for atom $\mu$. The filter-diagonalization technique was then applied to obtain quasiparticle electron states $\psi_i(\mathbf{r})$ above the conduction edge.  The calculations were implemented on real-space grids less than 0.8 a.u. such that the eigenenergies converges less than $10^{-3}$ meV. 

\begin{table}[ht]
\begin{tabular}{|c|c|c|c|c|}
\hline
   & $a_1$       & $a_2$     & $a_3$      & $a_4$     \\ \hline
Cd & -39.5761 & 1.0205 & -0.1361 & 1.6688 \\ \hline
Se & 5.7389   & 4.3792 & 1.2123  & 0.3197 \\ \hline
S  & 5.4875   & 4.3685 & 1.2135  & 0.2915 \\ \hline
\end{tabular}
\caption{Pseudopotential parameters for Cd, Se, and S. All parameters are
given in atomic units.
}
\label{tab:3}
\end{table}

To transform the delocalized quasi-electron eigenstates $\psi_i(\mathbf{r})$ to localized donor and acceptor states $\phi_n(\mathbf{r})$, the F\"{o}rster-Boys localization scheme is applied to maximize the self-extension criteria \cite{foster_canonical_1960,kleier_localized_1974}
\begin{equation*}
    \left< \hat{\Omega}\right>_{FB} = \sum_{n\in\mathcal{D},\mathcal{A}}\left(\int \mathrm{d}^3\mathbf{r} |\phi_n(\mathbf{r})|^2\mathbf{r}\right)^2,
\end{equation*}
where the optimization can be achieved by successive $2\times 2$ rotations of wavefunction pairs. The resulting localized states are related to eigenstates by a unitary matrix $U$
\begin{equation}
    \phi_n(\mathbf{r}) = \sum_{i}U_{ni}\psi_i(\mathbf{r}). 
\end{equation}
The first 8 eigenstates at the bottom of the conduction band will be included to construct localized states. 

\begin{figure}[ht]
    \centering
    {{\includegraphics[width=16cm]{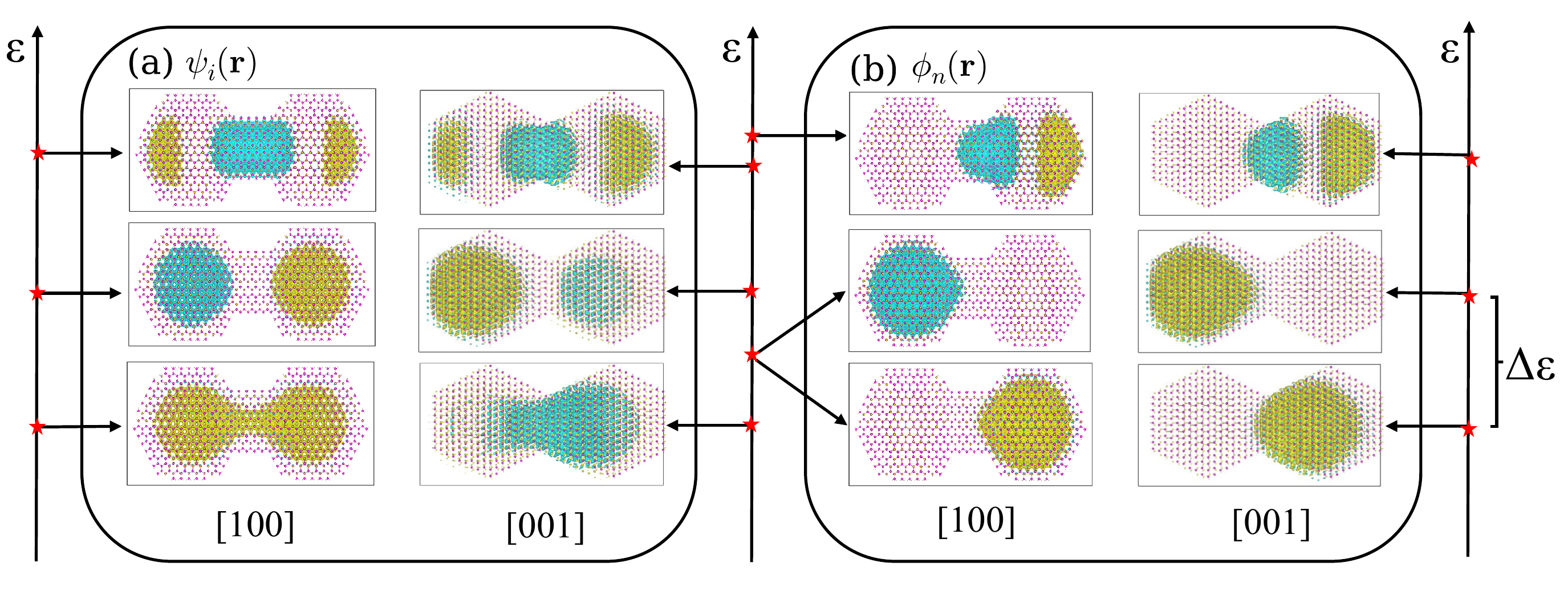}}}
    \caption{ Hyper-sphere plots of (a) eigenstates $\psi_i(\mathbf{r})$ and (b) local states $\phi_n(\mathbf{r})$ from the semi-empirical pseudopotential calculation and FB-localization of the quasi-electron eigenstates for wurtzite CQD dimers in symmetric [100] and asymmetric [001] attachment orientations. For the asymmetric orientation, there exists an energy bias $\Delta\varepsilon$ between the donor and acceptor states. }
    \label{fig:100-neck-core}
\end{figure}

\section{Normal Modes and Vibronic Couplings}
The nuclear vibration is described by the same 3-body Stillinger-Weber potential used in structure minimization. If harmonic approximation is made, the normal mode coordinates can be calculated by diagonalizing the Hessian matrix at the equilibrium configuration. For the mass-weighted coordinates, the Hessian matrix is 
\begin{equation}
    D_{\mu k,\mu' k'} = \frac{1}{\sqrt{m_\mu m_{\mu'}}}\left.\frac{\partial^{2} U_\text{SW}}{\partial u_{\mu k}\partial u_{\mu' k'}}\right|_{u=0},
\end{equation}
where the atomic displacement $u_{\mu k}=R_{\mu k}-R_{0,\mu k}, (k=x,y,z)$. The system-bath couplings in the atomic coordinates $V_{nm}^{\mu k}$ are given by the first-order derivative of pseudopotential with respect to the nuclear coordinates, \cite{jasrasaria_interplay_2021} and can be transformed to couplings to normal mode $V_{nm}^{\alpha}$ accordingly
\begin{align}
    V_{nm}^{\mu k}
    &=\int d \mathbf{r} \phi_{n}^*(\mathbf{r}) \frac{\partial \nu_{\mu}\left(\left|\mathbf{r}-\mathbf{R}_{\mu}\right|\right)}{\partial R_{\mu k}} \phi_{n}(\mathbf{r})\\
    V_{nm}^\alpha &= \sum_{\mu,k}\frac{1}{\sqrt{m_\mu}} E_{\mu k,\alpha}V_{nm}^{\mu k},
\end{align}
where $E_{\mu k,\alpha}$ are the coefficients from the normal mode transformation. 

\section{Ehrenfest Dynamics}
The dynamics of charge transfer is described by the mixed quantum-classical mean-field Ehrenfest method. The system dynamics can be derived from quantum-classical Liouville equation\cite{berkelbach_reduced_2012,kapral_progress_2006}
\begin{equation}\label{eq:ehren1}
\begin{aligned}
\frac{\partial \rho_{S}(t)}{\partial t} 
&=-\frac{i}{\hbar}\left[H_{S}+\sum_{\alpha}\sum_{\substack{n,m\in \mathcal{D},\mathcal{A}}} \ket{\phi_n}\bra{\phi_m}V_{nm}^\alpha{Q}_{\alpha}(t), \rho_{S}(t)\right],
\end{aligned}
\end{equation}
where $\rho_S(t)$ is the reduced density matrix. If we make harmonic approximation, then the time evolution of phonon coordinates is 
\begin{align}\label{eq:ehren2}
    \frac{dQ_\alpha(t)}{dt} &= P_\alpha(t)\\
    \label{eq:ehren3}
\frac{d P_{\alpha}(t)}{d t}
&=-\omega_\alpha^2Q_\alpha
-\sum_{\substack{n,m\in \mathcal{D},\mathcal{A}}} V_{nm}^\alpha \operatorname{Tr}_{S}\left\{\ket{\phi_n}\bra{\phi_m} \rho_{S}(t)\right\}.
\end{align}
The above equations of motion can be integrated by the fourth order Runge Kutta method with a time step around $1\sim3$ fs. The initial density matrix is assumed to be separated into a product of equilibrated system and bath operators
\begin{equation}
    \rho(0) =\rho_S(0)\rho_B(0)= \left(\sum_{n\in D}\frac{e^{-\beta E_n}}{Z_D}\ket{n}\bra{n}\right)\frac{e^{-\beta H_B}}{Z_B},
    \label{eq:rho_init}
\end{equation}
where $Z_D$ and $Z_B$ are partition functions for donor subspace and thermal bath. The initial bath coordinates $P_\alpha$ and $Q_\alpha$ are sampled from the Boltzmann distribution and the populations $\rho_S(t)$ are averaged over the ensemble. The calculation requires an average of $1000\sim4000$ trajectories to converge. The anharmonicity effects can be included by propagating the atomic coordinates in the force field of Stillinger–Weber potential. We found that the anharmonic bath does not change the dynamics significantly, and the comparison between the two dynamics are included in the following sections. 

The population dynamics, i.e. the time evolution of diagonal elements of $\rho_S(t)$ describes the charge transfer process. The population transfer between donor and acceptor subspaces can be described by $p_\mathcal{D}(t)$ and $p_\mathcal{A}(t)$ which are obtained by summing over all populations of donor and acceptor states respectively
\begin{equation}
    p_{\mathcal{D}/\mathcal{A}}(t) = \sum_{n\in\mathcal{D}/\mathcal{A}}\rho_{S,nn}(t).
\end{equation}
In general, there are two important time scales associated with charge transfer dynamics, and it is possible to express the envelop function $p_{\text{env}}(t)$ of $p_\mathcal{D}(t)$ (or $p_\mathcal{A}(t)$) as a sum of exponential functions:
\begin{equation}\label{eq:f_env}
p_{\text {env }}(t)= A_\text{M} e^{-k_\text{M}t} + A_{\rm dp} e^{-k_{\rm dp}t},
\end{equation}
where $k_\text{M}$ is Marcus rate, $k_{\rm dp}$ is the dephasing rate, and $A$ parameters are weighted factors. We choose $A_{\rm dp} = 0$ in the nonadiabatic Marcus regime and $A_{\rm M}=0$ in the adiabatic coherent regime. The total transfer rate $k$ is chosen as either $k_{\rm M}$ in the nonadiabatic regime or $k_{\rm dp}$ in the adiabatic regime.

\section{Population dynamics for individual states}
\begin{figure}[ht]
    \centering
    {{\includegraphics[width=16cm]{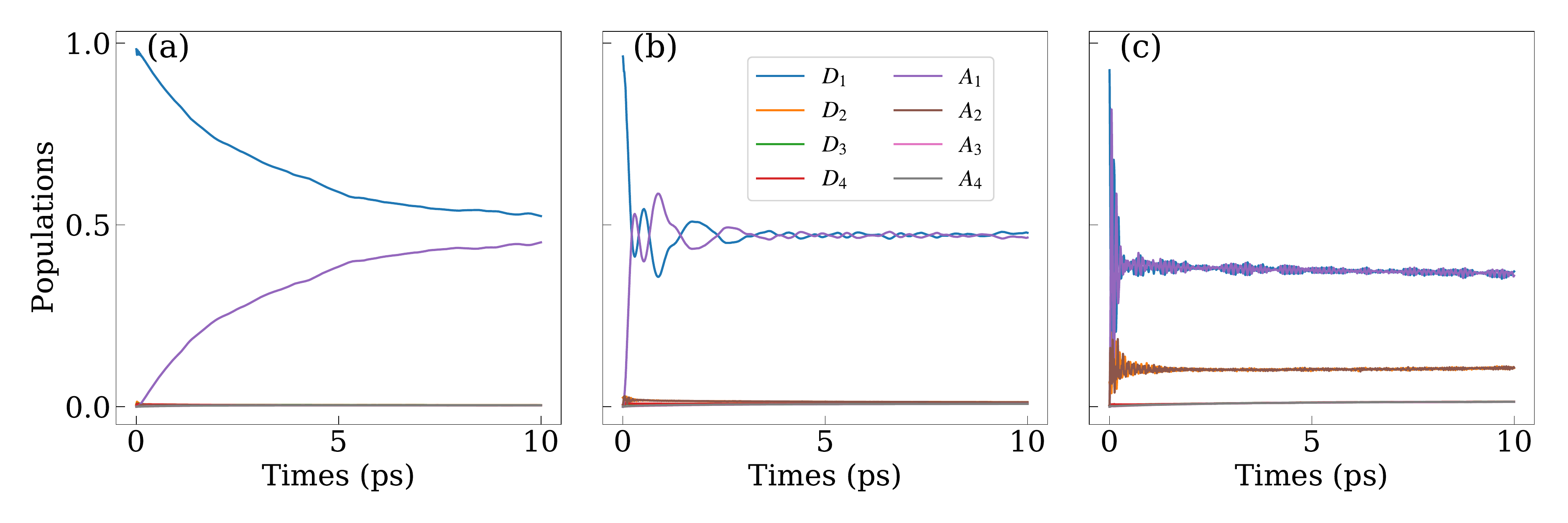}}}
    \caption{Complete population dynamics for 8 states (4 donor states and 4 acceptor states) of the systems shown in Fig. 3 in the main text. }
    \label{fig:pop-states}
\end{figure}

As discussed in the main text, 4 donor states and 4 acceptor states are taken to construct the Hamiltonian. The population dynamics for individual states corresponding to Fig.~3 in the main text are shown in Fig.~\ref{fig:pop-states}. The populations dynamics in Fig.~\ref{fig:pop-states} panel (a) and (b) are dominated by the two $1S_e$-like ground states $D_1$ and $A_1$. This is because the energy gap between $1S_e$ and $1P_e$ are much greater than the thermal energy in these two cases so $1P_e$-like states are not populated according to the initial condition Eq. \ref{eq:rho_init}. On the other hand, the $1S_e$-$1P_e$ energy gap is small enough in the case of Fig.~\ref{fig:pop-states} (c) such that $D_2$ is partially populated. The population transfer between $1S_e$ and $1P_e$-like states will thus influence to the total dynamics, such as the dephasing rate in the ground states.

\section{Population dynamics for all structures}
Fig.~\ref{fig:100-neck-core} to \ref{fig:001-mono-core} plots the population dynamics of all the CQD dimers with various $D_{\rm neck}$, $D_{\rm core}$ and $D_{\rm shell}$ studied in the main text. The label for each structure can be found in Table.~\ref{tab:1} and \ref{tab:2}. $\Delta \varepsilon$, $J$ and $\lambda$ are the energy bias, hybridization energy, and reorganization energy between two ground states (two $1S_e$-like $D_1$ and $A_1$ states). All calculations are performed at room temperature with $T=300$ K.

\begin{figure}[ht]
    \centering
    {{\includegraphics[width=16cm]{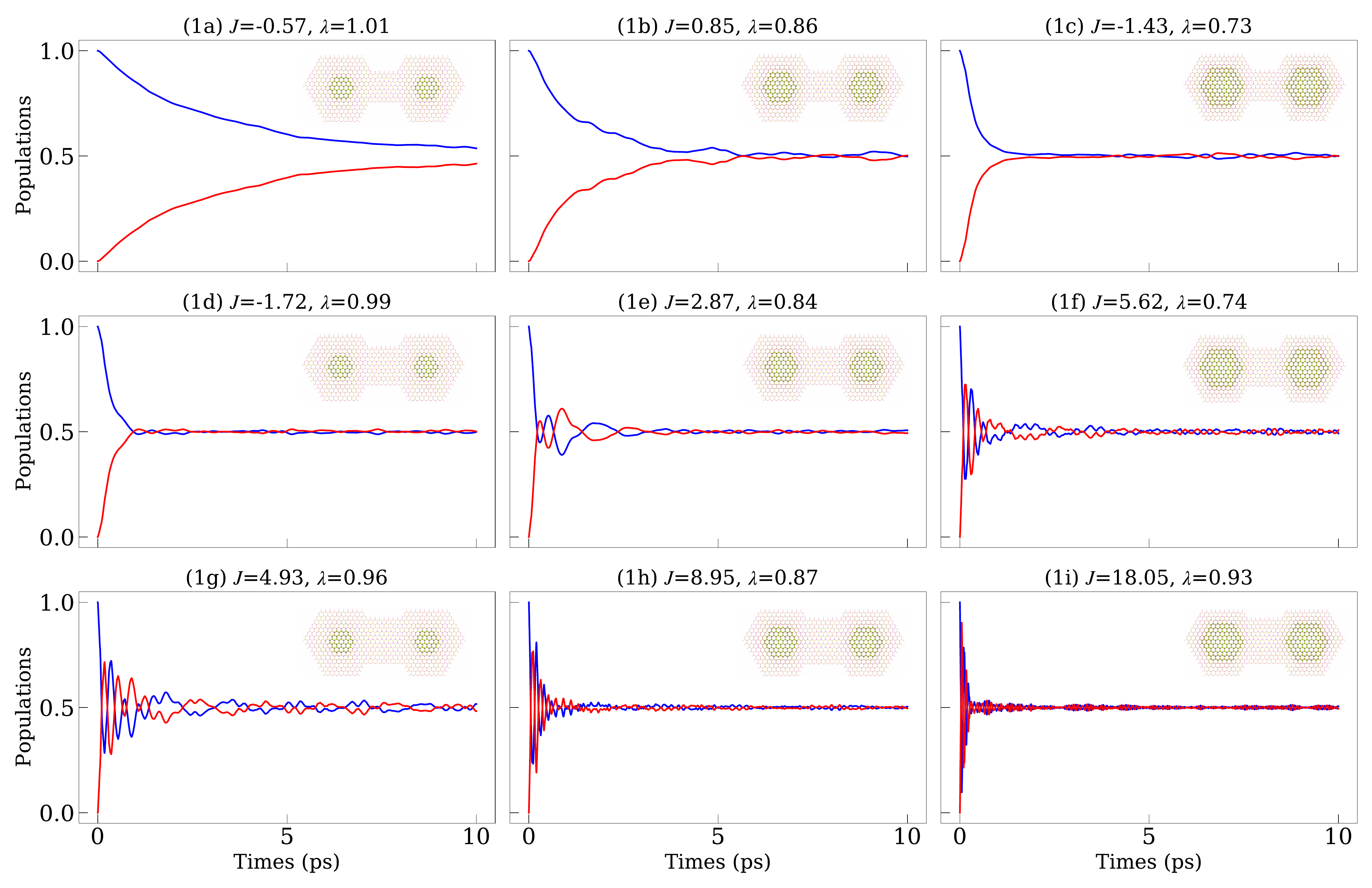}}}
    \caption{Population dynamics of donor $p_\mathcal{D}(t)$ (blue) and acceptor $p_\mathcal{A}(t)$ (red) at 300 K for different core diameters $D_{\rm core}$ (columns: 2.2, 3.0, 3.9 nm) and neck width $D_{\rm neck}$ (rows: 2.4, 3.1, 3.8 nm) in [100] attachment orientation. The unit for all parameters is in meV. The comparison between population dynamics shown in the main text Fig. 3 are taken from panels (1a), (1e) and (1i).}
    \label{fig:100-neck-core}
\end{figure}

\begin{figure}[ht]
    \centering
    {{\includegraphics[width=16cm]{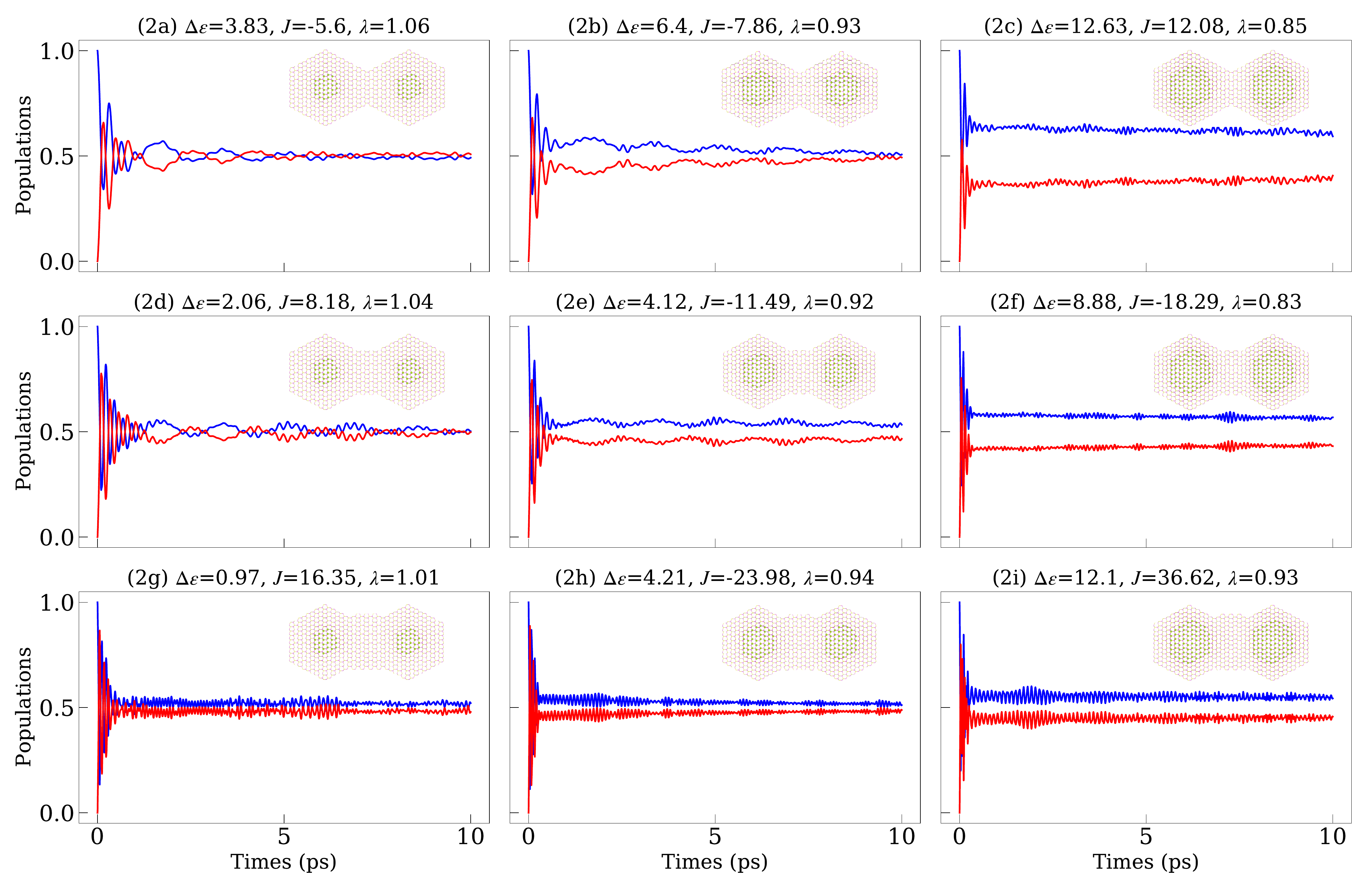}}}
    \caption{Population dynamics of donor $p_\mathcal{D}(t)$ (blue) and acceptor $p_\mathcal{A}(t)$ (red) at 300 K for different core diameters $D_{\rm core}$ (columns: 2.2, 3.0, 3.9 nm) and neck width $D_{\rm neck}$ (rows: 2.8, 3.5, 4.2 nm) in [001] attachment orientation. The unit for all parameters is in meV. }
    \label{fig:001-neck-core}
\end{figure}

\begin{figure}[ht]
    \centering
    {{\includegraphics[width=15cm]{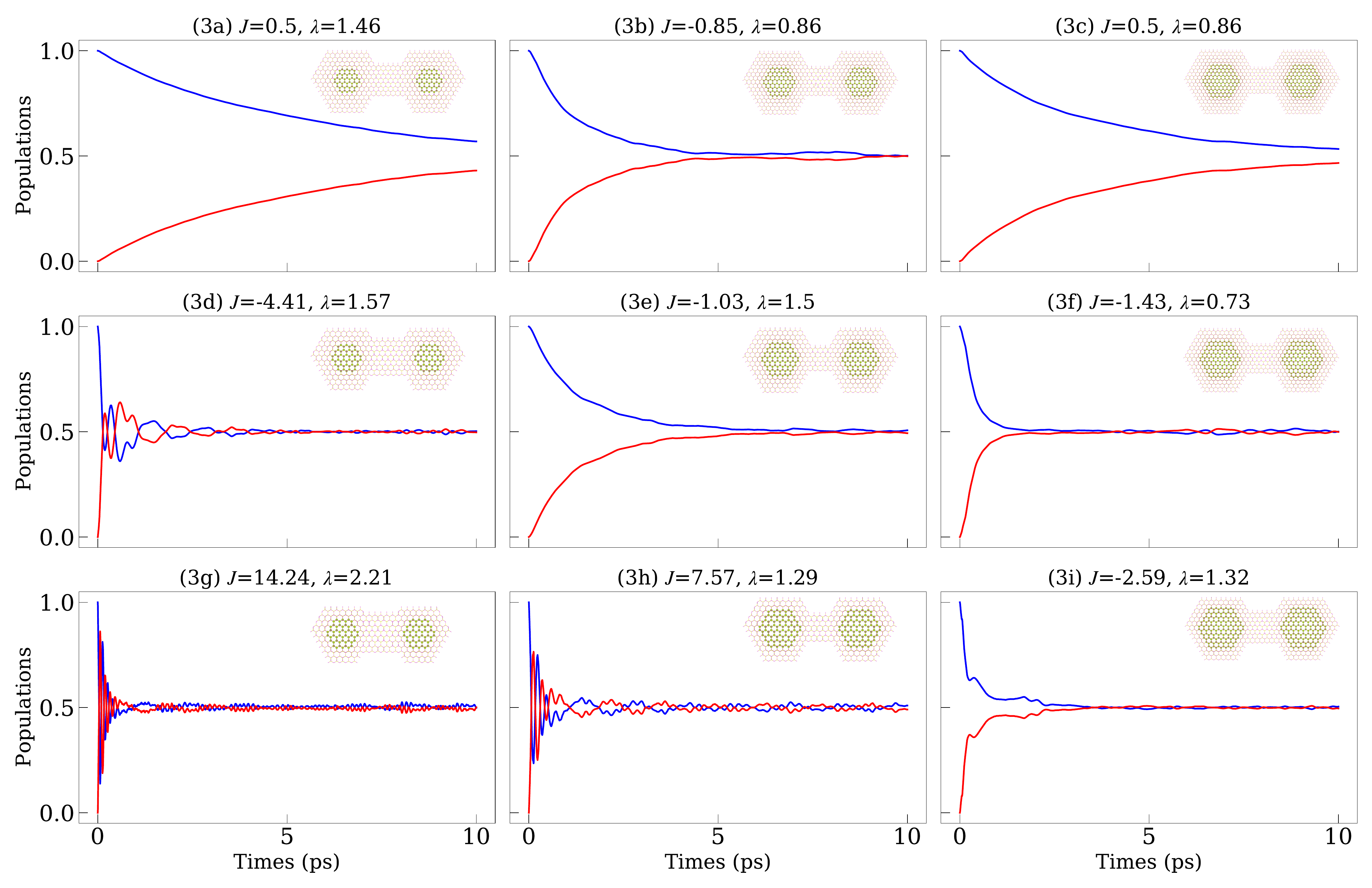}}}
    \caption{Population dynamics of donor $p_\mathcal{D}(t)$ (blue) and acceptor $p_\mathcal{A}(t)$ (red) at 300 K for different core diameters $D_{\rm core}$ (columns: 2.2, 3.0, 3.9nm) and shell thicknesses $D_{\rm shell}$ (rows: 5-layer, 4-layer, 3-layer) with [100] attachment orientation.}
    \label{fig:100-mono-core}
\end{figure}

\begin{figure}[ht]
    \centering
    {{\includegraphics[width=15cm]{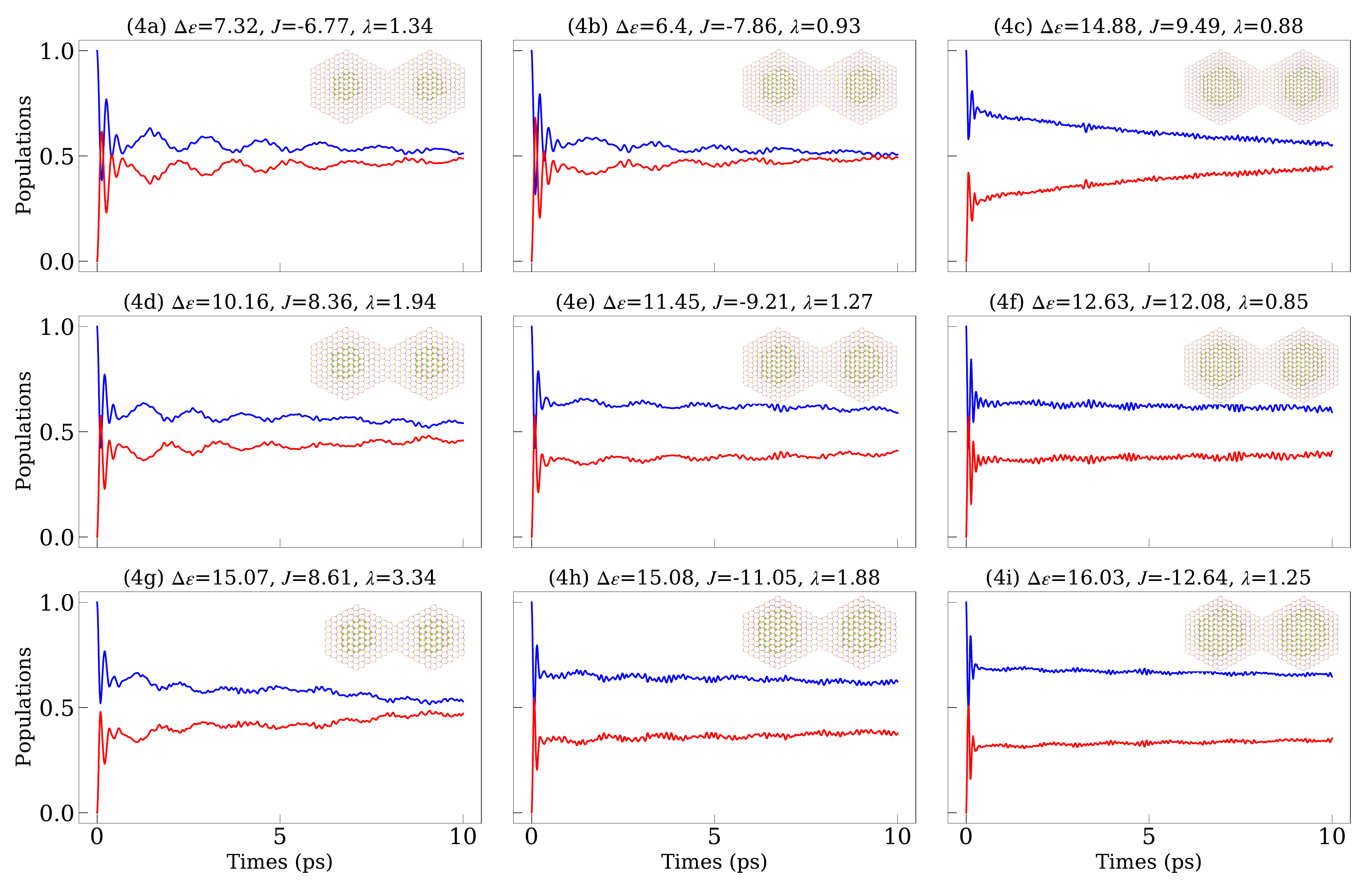}}}
    \caption{Population dynamics of donor $p_\mathcal{D}(t)$ (blue) and acceptor $p_\mathcal{A}(t)$ (red) at 300 K for different core diameters $D_{\rm core}$ (columns: 2.2, 3.0, 3.9nm) and shell  thicknesses $D_{\rm shell}$ (rows: 5-layer, 4-layer, 3-layer) with [001] attachment orientation.}
    \label{fig:001-mono-core}
\end{figure}

\clearpage
\section{Effects of off-diagonal couplings}
\begin{figure}[ht]
    \centering
    {{\includegraphics[width=15cm]{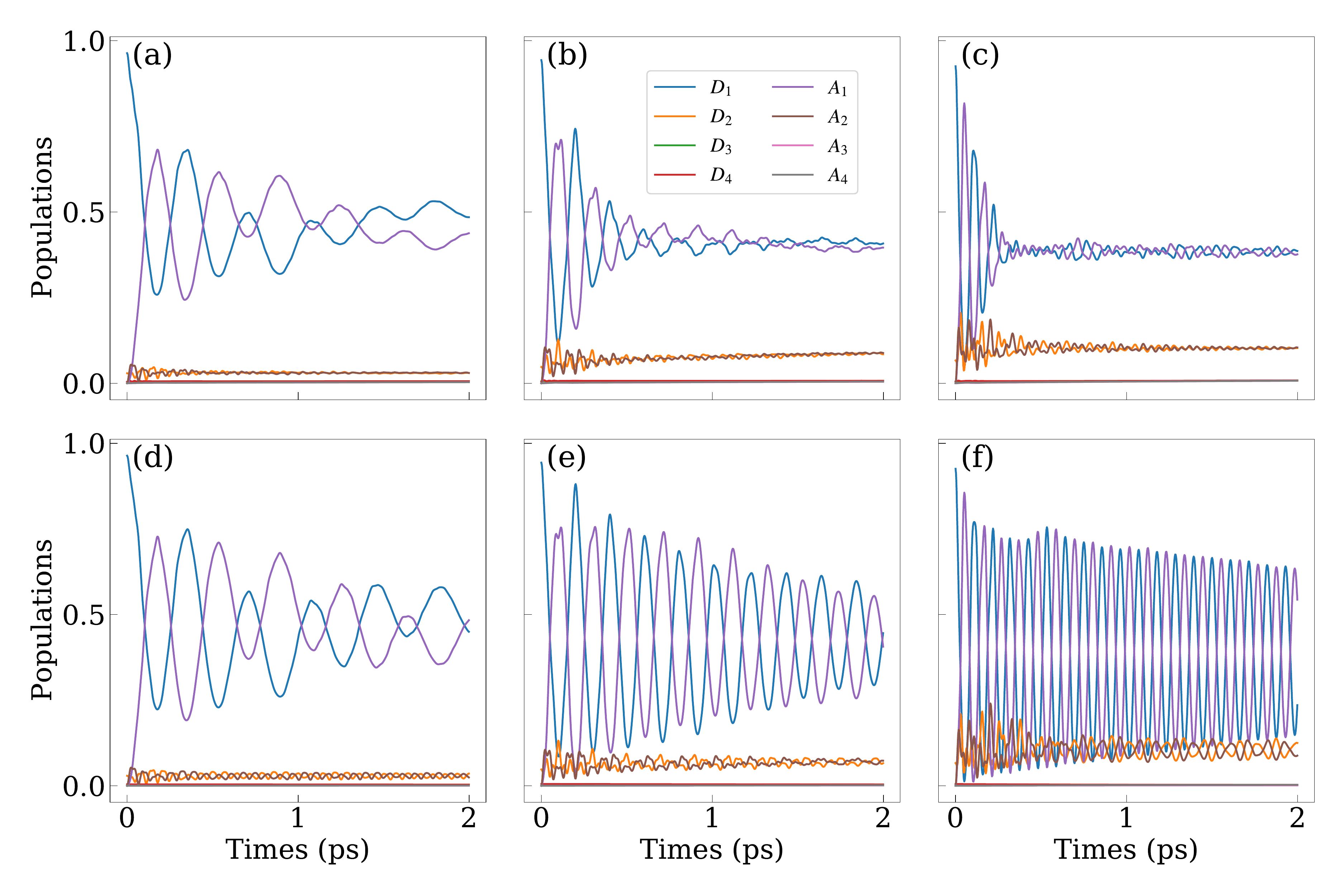}}}
    \caption{(a-c) Population of individual states for (1g), (1h) and (1i) respectively. Both diagonal couplings $V_{nn}^\alpha$ and off-diagonal couplings $V_{nm}^\alpha$ are included. (d-f) Population of (a-c) with off-diagonal couplings removed $V_{nm}^\alpha$ ($n,m\in\mathcal{D},\mathcal{A}$ and $n\neq m$). }
    \label{fig:off-diag-cp}
\end{figure}

In the main text, we argued that the off-diagonal couplings along with population transfer between $1S_e$-like ($D_1, A_1$) and $1P_e$-like ($D_2, A_2$) states are responsible for the dephasing trend seen in the adiabatic regime. Fig.~\ref{fig:off-diag-cp} show comparison between three adiabatic dynamics when including both diagonal and off-diagonal vibronic couplings (a-c) and dynamics when only diagonal couplings are included (d-f). Note that the decreasing rate decrease significantly when the off-diagonal couplings are removed yielding longer population coherence. 

\clearpage
\bibliography{supplement}